\documentclass{aa}

\usepackage{amsmath,array}

\usepackage{txfonts}
\usepackage{sansmath}

\usepackage{dcolumn}
\newcolumntype{d}[1]{D{.}{.}{#1} }

\usepackage[usenames]{color}
\definecolor{darkgreen}{rgb}{0.,0.7,0.}
\usepackage{ulem}
\renewcommand{\emph}[1]{\textit{#1}}

\begin{document}

\title{The Galactic Center Molecular Cloud Survey}
\subtitle{I. A Steep Linewidth--Size Relation \& Suppression of Star Formation}

\author{Jens~Kauffmann \inst{1}
  \and 
  Thushara~Pillai \inst{1}
  \and
  Qizhou~Zhang \inst{2}
  \and
  Karl~M.~Menten \inst{1}
  \and
  Paul~F.~Goldsmith \inst{3}
  \and
  Xing~Lu \inst{2,4}
  \and
  Andr\'es~E.~Guzm\'an \inst{5}
}

\institute{Max--Planck--Institut f\"ur Radioastronomie, Auf dem
  H\"ugel 69, 53121 Bonn, Germany
  \and
  Harvard--Smithsonian Center for Astrophysics, 60 Garden Street,
  Cambridge, MA 02138, USA
  \and
  Jet Propulsion Laboratory, California Institute of Technology, 4800
  Oak Grove Drive, Pasadena, CA 91109, USA
  \and
  School of Astronomy and Space Science, Nanjing University, 22 Hankou
  Road, Nanjing 210093, China
  \and
  Departamento de Astronom\'ia, Universidad de Chile, Camino el
  Observatorio 1515, Las Condes, Santiago, Chile}

\abstract{The Central Molecular Zone (CMZ; inner $\sim{}200~\rm{}pc$)
  of the Milky Way is a star formation (SF) environment with very
  extreme physical properties
  \citep{morris1996:cmz-review}. Exploration of SF in this region is
  important because (\textit{i})~this region allows us to test models
  of star formation under exceptional conditions, and
  (\textit{ii})~the CMZ clouds might be suitable to serve as templates
  to understand the physics of starburst galaxies in the nearby and
  the distant universe. For this reason we launched the Galactic
  Center Molecular Cloud Survey (GCMS), the first systematic study
  that resolves all major CMZ clouds at interferometer angular
  resolution (i.e., a few arc seconds). Here we present initial
  results based on observations with the Submillimeter Array (SMA) and
  the Atacama Pathfinder Experiment (APEX). Our study is complemented
  by dust emission data from the Herschel Space Telescope and a
  comprehensive literature survey of CMZ star formation activity. Our
  research reveals (\textit{i})~an unusually steep linewidth--size
  relation, $\sigma(v)\propto{}r_{\rm{}eff}^{0.66\pm{}0.18}$, down to
  velocity dispersions $\sim{}0.6~\rm{}km\,s^{-1}$ at 0.1~pc scale.
  This scaling law potentially results from the decay of gas motions
  to transonic velocities in strong shocks. The data also show that,
  relative to dense gas in the solar neighborhood, (\textit{ii})~star
  formation is suppressed by factors $\gtrsim{}10$ in
  \emph{individual} CMZ clouds. This observation encourages
  exploration of processes that can suppress SF \emph{inside} dense
  clouds for a significant period of time.}


\maketitle


\defcitealias{kauffmann2016:gcms_ii}{Paper~II}

\section{Introduction\label{sec:introduction}}
The Central Molecular Zone (CMZ) --- the inner $\sim{}200~\rm{}pc$ of
our Galaxy --- is a star--forming environment with extreme physical
properties. About 3--10\% of the total molecular gas and star
formation (SF) of the Milky Way\footnote{Observations of line and dust
  emission indicate that the $|\ell|\le{}3\degr$ region contains of
  order $(3~{\rm{}to}~8)\times{}10^7\,M_{\sun}$ of molecular gas
  \citep{dahmen1998:cmz-gas, tsuboi1999:cs-nobeyama,
    longmore2012:sfr-cmz}. It forms stars at a rate of order
  $0.1\,M_{\sun}\,\rm{}yr^{-1}$ \citep{yusef-zadeh2009:sf-cmz,
    immer2012:recent-sfr, longmore2012:sfr-cmz,
    koepferl2015:masquerading-cmz}. The Milky Way contains
  $(1.0\pm{}0.3)\times{}10^9\,M_{\sun}$ of molecular gas
  \citep{heyer2015:review}, while the star formation rate is
  $1~{\rm{}to}~4\,M_{\sun}\,\rm{}yr^{-1}$ \citep{diehl2006:26al,
    misiriotis2006:ism-3d, lee2012:wmap}.} reside at
$|\ell|\le{}3\degr$ (i.e., within galactocentric radii
$\le{}430~\rm{}pc$). CMZ molecular clouds have unusually high mean
$\rm{}H_2$ densities $\sim{}10^4~\rm{}cm^{-3}$ and column densities
$\sim{}10^{23}~\rm{}cm^{-2}$ (e.g., \citealt{lis1994:dust-ridge}), and
they are subject to an average gas pressure from hot gas of order
$10^{6~{\rm{}to}~7}~\rm{}K\,cm^{-3}$ \citep{yamauchi1990:cmz-plasma,
  spergel1992:pressure-bulge, muno2004:diffuse-xray}. The gas is
pervaded by a strong magnetic field of a few $10^3~\rm{}\mu{}G$
\citep{yusef-zadeh1984:non-thermal,
  uchida1985:cmz-lobes,chuss2003:cmz-polarization,
  novak2003:cmz-polarization} that also penetrates the CMZ clouds
\citep{pillai2015:magnetic-fields}. Unusually wide lines
$\gtrsim{}10~\rm{}km\,s^{-1}$ on spatial scales $\gtrsim{}1~\rm{}pc$
(e.g., \citealt{guesten1981:nh3-cmz}, strong and widespread SiO
emission tracing shocks \citep{martin-pintado1997:sio-cmz,
  huettemeister1998:shocks, riquelme2010:survey}, prevalence of
molecules likely ejected from grain surfaces via shocks
\citep{requena-torres2006:corganic-mols,
  requena-torres2008:coxygen-coms}, and collisionally--excited
methanol masers 
(\citealt{mills2015:cmz-masers}; also see \citealt{menten2009:g1.6},
though) suggest that much of the gas in the CMZ is subject to violent
gas motions, such as cloud--cloud collisions at high velocities. Gas
temperatures are typically in the range $50~{\rm{}to}~100~\rm{}K$
(\citealt{guesten1981:nh3-cmz, huettemeister1993:nh3-cmz,
  ao2013:cmz-temperatures, mills2013:widespread-hot-nh3,
  ott2014:cmz-atca, ginsburg2015:cmz-gas-temperatures}; also see
\citealt{riquelme2010:isotopes,
  riquelme2012:temp-loop-interceptions}). In stark contrast to
  spiral--arm clouds, and despite relatively high gas densities, the
high gas temperature in the CMZ is mysteriously decoupled
from the much lower temperature of dust grains, for which values
$\sim{}20~\rm{}K$ are obtained (e.g., \citealt{guesten1981:nh3-cmz},
\citealt{molinari2011:cmz-ring}, \citealt{longmore2011:m025}; see
\citealt{clark2013:gas-dust-g0253} for some modeling work). The CMZ is
populated by numerous exotic objects, such as masers that are possibly
enabled by the extreme physical conditions of the environment (e.g.,
\citealt{ginsburg2015:g0.38}). The clouds reside at a distance of
$8.34\pm{}0.16~\rm{}kpc$ (\citealt{reid2014:bessel}; this value is
adopted throughout this study), and they appear to lie on a
well--organized common orbit (e.g., \citealt{molinari2011:cmz-ring}
and \citealt{kruijssen2014:orbit}) along which clouds might also
systematically evolve \citep{longmore2013:cmz-cluster-progenitors}.

The study of CMZ molecular clouds is critical for two reasons. First,
the CMZ allows to explore an extreme point in the star formation
parameter space: valid models of SF physics must describe star
formation in the CMZ environment as well as under the conditions
prevailing in clouds closer to sun. Second, the CMZ might serve as a
template for unresolved processes that are active in nearby and more
distant starburst galaxies.

For this reason we launched the Galactic Center Molecular Cloud Survey
(GCMS), the first systematic study resolving all major CMZ molecular
clouds at interferometer angular resolution. This work is inspired by
first interferometric investigations of the CMZ cloud
G0.253+0.016\footnote{Observations of G0.253+0.016 with the Midcourse
  Space Experiment (MSX) inspired \citet{egan1998:irdcs} and
  \citet{carey1998:irdc-properties} to coin the now--popular term of
  the Infrared Dark Cloud (IRDC).}  (\citealt{guesten1981:nh3-cmz};
\citealt{lis1994:m0.25, lis2001:ir-spectra}; \citealt{lis1998:m025};
a.k.a.\ the ``Brick'': \citealt{longmore2011:m025}) with the
Submillimeter Array (SMA; \citealt{kauffmann2013:g0.253} and
\citealt{johnston2014:g0.253}) and ALMA
\citep{rathborne2014:g0253-pdf,
  rathborne2015:g0253-alma}. Section~\ref{sec:observations} introduces
the GCMS. Some of the observations were already explored by
\citet{kauffmann2013:g0.253}, \citet{kendrew2013:sgr-c}, and
\citet{lu2015:20kms}. Here we present a first study that characterizes
all targets in a homogeneous way and permits to draw general
conclusions about the structure of CMZ molecular clouds.

Our study employs observations of dust and the $\rm{}N_2H^+$~(3--2)
transition. We have made this choice because these probes are believed
to be faithful tracers of the densest gas in molecular clouds
(Sec.~\ref{sec:obs-sma-apex} summarizes properties of
$\rm{}N_2H^+$).\medskip

\noindent{}In this paper we focus on two particular topics. First, we
conduct a systematic exploration of the kinematics of CMZ molecular
clouds. In \citet{kauffmann2013:g0.253} we demonstrated for the first
time that line widths $\lesssim{}1~\rm{}km\,s^{-1}$ prevail in some CMZ cloud
fragments on spatial scales $\lesssim{}0.1~\rm{}pc$. This is
remarkable given the much larger aforementioned line widths
$\gtrsim{}10~\rm{}km\,s^{-1}$ on spatial scales
$\gtrsim{}1~\rm{}pc$. In essence this exploration of cloud kinematics
over a range of spatial scales constrains the
linewidth--size relation (e.g.,
\citealt{larson1981:linewidth_size}). While this concept has been explored extensively
for entire regular Milky Way clouds (starting with, e.g.,
\citealt{sanders1985:gmcs_ii} and \citealt{solomon1987:co-survey}),
relatively little work has been done on CMZ clouds (e.g.,
\citealt{oka1998:co-cmz_ii}, \citealt{miyazaki2000:dense-gas-cmz_ii},
and \citealt{shetty2012:linewidth-size-cmz}). These papers find that
the velocity dispersion scales with spatial scale as
$\sigma(v)\propto{}\ell^{h_{\sigma(v)}}$, where the linewidth--size
slope, $h_{\sigma(v)}$, lies in the range 0.4 to 0.8, depending on the
author, method, and emission line studied. The studies agree, though,
that $h_{\sigma(v)}$ appears to be the same inside and outside the
CMZ. Here we come to a different conclusion because of the small line
widths now revealed by the interferometer data on spatial scales
$\lesssim{}0.1~\rm{}pc$.

Second, we explore the star formation rates of individual molecular
clouds in the CMZ. One of the surprising trends in the CMZ is that,
despite the high average gas densities, the star formation rate per
unit dense gas is suppressed when compared to the solar
neighborhood. For example, \citet{guesten1983:h2o-masers} and
\citet{caswell1983:water-masers} found early on that very few
$\rm{}H_2O$ masers reside in the CMZ, relative to the high mass of
dense gas. \citet{caswell1996:masers-methanol} later showed the same
for class~II methanol masers that uniquely trace high--mass star
formation. \citet{taylor1993:cmz-water-masers} concluded that, based
on the mass and density of molecular material, the CMZ should contain
about an order of magnitude more $\rm{}H_2O$ masers than
observed. Observations of the dense gas with increased angular
resolution at wavelengths $\lambda{}\lesssim{}1~\rm{}mm$ eventually
revealed individual clouds with little star formation
(\citealt{lis1994:m0.25, lis2001:ir-spectra};
\citealt{lis1998:m025}). The most recent studies established that SF
relations that apply in the solar neighborhood
\citep{heiderman2010:sf-law, lada2010:sf-efficiency,
  evans2014:sfr-nearby-clouds} overpredict CMZ star formation by an
order of magnitude, both on the scale of the entire CMZ
\citep{longmore2012:sfr-cmz} and on that of individual clouds
\citep{kauffmann2013:g0.253}. Here we expand this research by
obtaining individual star formation rates for all major CMZ molecular
clouds. One interesting hypothesis is that CMZ star formation in the
dense gas is suppressed because \emph{all} CMZ molecular clouds are
young and are in a state that precedes efficient star formation. Our
new data on individual clouds allow to test and reject this scenario.

A number of aspects can be explored with our data and the GCMS results
will be published in a series of papers. We therefore also include a
broad description of the data set in this initial paper. For example,
a companion study published simultaneously with the current work ---
\citet{kauffmann2016:gcms_ii}, hereafter \citetalias{kauffmann2016:gcms_ii} --- explores the
density structure of CMZ clouds and explores its relationship with the
observed star formation activity.\medskip

\noindent{}The current paper is organized as
follows. Section~\ref{sec:observations} describes the sample, the
observations, and the data reduction. In Sec.~\ref{sec:kinematics} we
characterize and analyze the kinematic structure of CMZ molecular
clouds. The star formation characteristics of the clouds are then
discussed in Sec.~\ref{sec:suppression-sf}. General conclusions on the
structure of CMZ molecular clouds are drawn in
Sec.~\ref{sec:discussion}. A summary of the study is provided in
Sec.~\ref{sec:summary}. The work in the paper is supported by
appendices on imaging of interferometer data
(Appendix~\ref{sec-app:imaging}), observations of CMZ star formation
(Appendix~\ref{sec-app:sf-observations}), and methods to determine the
CMZ star formation rate (Appendix~\ref{sec-app:sf-rate}).

\section{Sample, Observations \& Data
  Reduction\label{sec:observations}}
\subsection{Sample Selection}
\begin{figure*}
\includegraphics[width=\linewidth]{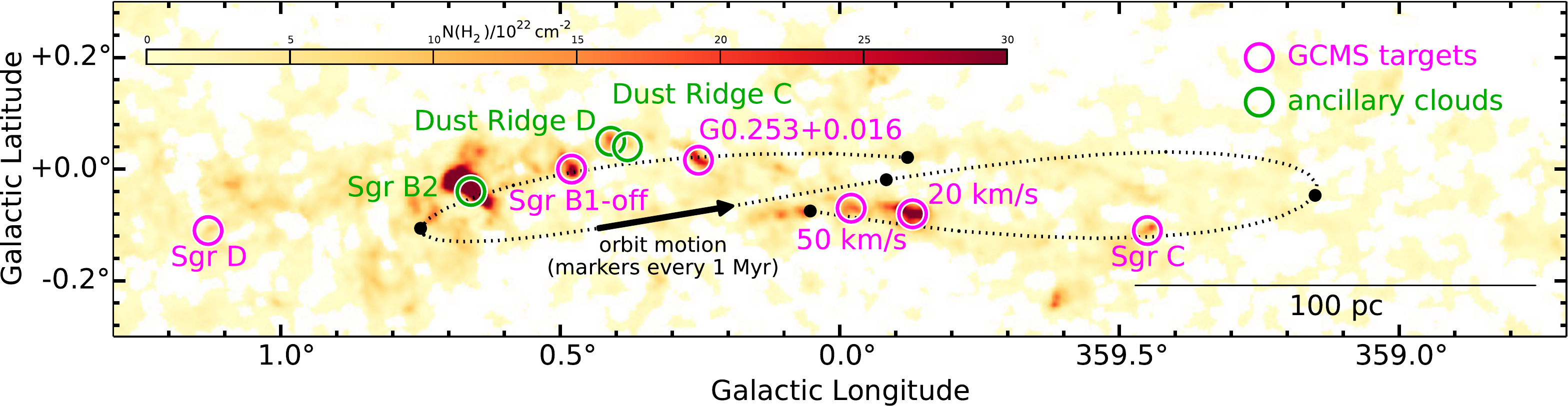}
\caption{Overview of the inner CMZ. Clouds targeted by GCMS
  interferometer observations presented here are highlighted in
  \emph{magenta}. Other ancillary clouds discussed in this paper are
  shown in \emph{green}. A \emph{dotted line} shows part of the orbit
  for CMZ clouds proposed by \citet{kruijssen2014:orbit}. The
  \emph{background image} presents a column density map derived from
  Herschel dust emission data, as described in this
  paper.\label{fig:overview}}
\end{figure*}

Previous research into G0.253+0.016 demonstrates that investigating
massive and dense CMZ molecular clouds informs both general studies of
star formation and research into the processes acting in the CMZ. The
GCMS is therefore designed to provide a general overview of the
physical conditions in dense and massive CMZ molecular clouds
resembling G0.253+0.016. Figure~\ref{fig:overview} and
Table~\ref{tab:target-summary} give an overview of the regions studied
here, including some ancillary clouds which we characterize using
archival data. The masses and radii listed in
Table~\ref{tab:target-summary} are derived in
Sec.~\ref{sec:obs-herschel} using Herschel data.  We select target
regions on the basis of SCUBA $850~\rm{}\mu{}m$ dust emission maps
\citep{pierce-price2000:cmz_scuba} collected from the SCUBA Legacy
Catalogue
archive\footnote{\url{http://www2.cadc-ccda.hia-iha.nrc-cnrc.gc.ca/community/scubalegacy/}}
(\citealt{difrancesco2008:scuba_catalogue}; the beam size is
$19\arcsec$ at $850~\rm{}\mu{}m$ wavelength). Manual experimentation
reveals that the SCUBA $850~\rm{}\mu{}m$ contour at
$5~\rm{}Jy\,beam^{-1}$ traces the outline of G0.253+0.016 very
well. Further targets are therefore selected by searching the maps for
clouds exceeding the $5~\rm{}Jy\,beam^{-1}$ intensity
threshold. Assuming emission from dust at a temperature of $20~\rm{}K$
and opacities following \citet{ossenkopf1994:opacities}, this contour
corresponds to an $\rm{}H_2$ column density
$\sim{}2\times{}10^{23}~\rm{}cm^{-2}$
\citep{kauffmann2008:mambo-spitzer}.

\begin{table*}
  \caption{Basic information on the clouds discussed in this
    paper. Radius and mass refer to the cloud area above an $\rm{}H_2$
    column density of $9\times{}10^{22}~\rm{}cm^{-2}$, as derived from
    Herschel data in 
    Sec.~\ref{sec:obs-herschel}. The second to last column refers to the maximum
    mass per Herschel beam of $37\arcsec$ (i.e.,
    $r_{\rm{}eff}=0.74~\rm{}pc$) found within the aforementioned
    contour.\label{tab:target-summary}}
\begin{center}
\begin{tabular}{llld{1}ld{1}lllllllllllll}
\hline \hline
Target & Location $(\ell, b)$ & Radius & \multicolumn{2}{l}{Mass} &
 \multicolumn{2}{l}{Peak Mass per Beam} & Mean $\rm{}H_2$ Particle Density\\
 & & pc & \multicolumn{2}{l}{$10^4\,M_{\sun}$} &
 \multicolumn{2}{l}{$10^3\,M_{\sun}$} & $10^4~\rm{}cm^{-3}$\\ \hline
\multicolumn{7}{l}{\itshape{}observed in this paper:}\\
Sgr~C & $359\fdg{}45,-0\fdg{}11$ & 1.7 &
2.5 & & 7.9 & & 1.8\\
$20~\rm{}km\,s^{-1}$ & $359\fdg{}87,-0\fdg{}08$ & 5.1 &
33.9 & & 17.6 & & 0.9\\
$50~\rm{}km\,s^{-1}$ & $359\fdg{}98,-0\fdg{}07$ & 2.7 &
6.5 & & 7.2 & & 1.2\\
G0.253+0.016 & $0\fdg{}25,+0\fdg{}02$ & 2.8 &
9.3 &&  10.3 & & 1.5\\
Sgr~B1--off & $0\fdg{}48,+0\fdg{}00$ & 3.6 &
14.5 & & 15.5 & & 1.1\\
Sgr~D & $1\fdg{}13,-0\fdg{}11$ & --- &
\multicolumn{1}{r}{---} & & 3.4 & & ---\\ \\
\multicolumn{7}{l}{\itshape{}ancillary CMZ clouds:}\\
Dust Ridge C & $0\fdg{}38,+0.04$ & --- & \multicolumn{1}{r}{---} & &
3.9 & & ---\\
Dust Ridge D & $0\fdg{}41,+0.05$ & 2.0 &
3.7 & & 7.7 & & 1.6\\
Sgr~B2 & $0\fdg{}66,-0.04$ & 7.1 &
136.5 & & 101.8 & & 1.3\\
\hline
\end{tabular}
\end{center}
\end{table*}

We limit our study to the region between Sgr~C and Sgr~B2. Target
regions are further required to be large enough to sample a
significant mass reservoir. This selection is done manually by visual
inspection of the target sizes. The least massive region in our sample
contains a mass of $2.5\times{}10^{4}\,M_{\sun}$. Our survey is
effectively complete to clouds exceeding this mass. We further exclude
regions too large to be mapped in reasonable time. Only one target,
the Sgr~B2 complex, is removed from our sample for this
reason. However, \citetalias{kauffmann2016:gcms_ii} uses data from
\citet{schmiedeke2016:sgrB2-dust} to fill in SMA--based density
information on Sgr~B2. In a final step we manually add the Sgr~D
region to this data set. This region was observed in the same setup
used for other targets. However, the target resides slightly outside
the region considered here.

We caution that Sgr~D is likely to be a foreground or background
object that is not physically associated with the CMZ. First, this
region has relatively narrow lines (e.g.,
\citealt{mehringer1998:sgr_d}; also see our results below). Second,
\citet{mehringer1998:sgr_d} demonstrate that broad absorption lines
clearly associated with CMZ material are seen against the continuum of
the Sgr~D \ion{H}{ii}~region. This suggests that the
\ion{H}{ii}~region resides behind the CMZ as discussed here. Third,
\citet{blum1999:extinction-sgrd} infer the presence of a molecular
cloud extincting background stars at near--infrared wavelengths that
must reside in front of the CMZ. \citet{sawada2009:sgr_d} combine the
evidence to conclude that the entire Sgr~D complex consists of several
physically unrelated components that are by chance seen along the same
line of sight. For these reasons we follow a conservative approach
when using data for this region. \emph{We present information for
  Sgr~D as if located in the CMZ, but only as long as the region is
  easily identified in the Tables and Figures.} Our conclusions are
not influenced by this uncertainty.

This leaves us with a list of 6 targets. We here use observations of
these regions that were collected using the SMA and the Atacama
Pathfinder Experiment (APEX). \citet{kauffmann2013:g0.253} already
examined the SMA data for G0.253+0.016, and \citet{kendrew2013:sgr-c}
used some of the observations for the Sgr~C region. The SMA continuum
and $\rm{}N_2H^+$ data for the remaining clouds, plus the APEX data
for all targets, have not previously been presented.

\subsection{SMA \& APEX Observations\label{sec:obs-sma-apex}}
\paragraph{Interferometer Data} We used the Submillimeter Array
(SMA\footnote {The Submillimeter Array is a joint project between the
  Smithsonian Astrophysical Observatory and the Academia Sinica
  Institute of Astronomy and Astrophysics, and is funded by the
  Smithsonian Institution and the Academia Sinica.})  to observe the
$\rm{}N_2H^+$ (3--2) line and dust continuum near 280~GHz. An
observing log is provided in Table~\ref{tab:observations-sma}.  A
total of four tracks was obtained on 2009 June 2 and 4
($\approx{}0.44~\rm{}km\,s^{-1}$ resolution, $4~\rm{}GHz$ combined
total bandwidth) and 2012 May 29 and 30
($\approx{}0.87~\rm{}km\,s^{-1}$ resolution, $8~\rm{}GHz$ total
bandwidth). Every target was mosaiced by observing several positions
separated at less than half a $42\arcsec$ primary beam. The 345~GHz
receiver was tuned to the $\rm{}N_2H^+$ line in the LSB spectral band
s4, using 256 channels per chunk and 24 chunks per sideband in 2009
June, respectively 128 channels per chunk and 24 chunks per sideband
in 2012 May. This produces LSB frequency ranges of
$\nu/{\rm{}GHz}=[277.8,279.8]$ in 2009 and $[275.8,279.8]$ in 2012,
and USB ranges of $[287.8,289.8]$ and $[287.8,291.8]$,
respectively. The bandpass calibrator was 3C273 in 2009, while 3C379
was used in 2012. Flux calibration is based on Titan in 2009, and on
Uranus in 2012. Seven antennas were available in each observing
session. They were arranged in the ``compact north'' and ``compact''
configurations in 2009 and 2012, respectively. Data were taken under
good weather conditions with zenith optical depth $<0.1$ at 225~GHz
and single--sideband system temperatures $<200~\rm{}K$. Further
observations of G0.253+0.016 at about 219~GHz from 2012 May were taken
as part of this project, but these data are not relevant for the
current publication.

\begin{table*}
\caption{Observing log for the SMA observations exploited in this
  paper.\label{tab:observations-sma}}
\begin{center}
\begin{tabular}{llllllllllllllllll}
\hline \hline
Date & Targets & Setup & Flux of 1744$-$312\\ \hline
2009 June 2 & Sgr C, Sgr D & 4~GHz\tablefootmark{a}, $0.44~\rm{}km\,s^{-1}$ &
                       0.23~mJy\\
2009 June 4 & G0.253+0.016 & 4~GHz\tablefootmark{a}, $0.44~\rm{}km\,s^{-1}$ &
                       0.28~mJy\\
2012 May 29 & $50~\rm{}km\,s^{-1}$, Sgr B1 Off & 8~GHz\tablefootmark{b},
                       $0.87~\rm{}km\,s^{-1}$ &
                       0.23~mJy\\
2012 May 30 & $20~\rm{}km\,s^{-1}$ & 8~GHz\tablefootmark{b},
                       $0.87~\rm{}km\,s^{-1}$ &
                       0.20~mJy\\
\hline
\end{tabular}
\tablefoot{
\tablefoottext{a}{frequencies of
    $277.8~\text{to}~279.8~\rm{}GHz$ and
    $287.8~\text{to}~289.8~\rm{}GHz$}
\tablefoottext{b}{frequencies of
    $275.8~\text{to}~279.8~\rm{}GHz$ and
    $287.8~\text{to}~291.8~\rm{}GHz$}
}
\end{center}
\end{table*}

The data were calibrated within
MIR\footnote{\url{https://www.cfa.harvard.edu/sma/mir/}} using
standard procedures. This involves flagging as well as the calibration
of bandpass, flux, and phase. Consistency checks on the flux scale
were done on 1744$-$312. The flux of this source varies on long
timescales, but Table~\ref{tab:observations-sma} shows that the flux
calibration is consistent between consecutive observations.

The spectral data are further filtered using the UVLIN task from the
MIRIAD\footnote{\url{http://www.atnf.csiro.au/computing/software/miriad/}}
package \citep{sault2006:miriad}. Specifically, line--free channels
are used to produce an averaged continuum signal for continuum
imaging. The same channels are also used to estimate and subtract the
continuum signal for later imaging of pure line emission.

\paragraph{Single--Dish Data} Complementary single--dish observations
of the $\rm{}N_2H^+$ (3--2) line were obtained with the Atacama
Pathfinder Experiment (APEX\footnote{This publication is based on data
  acquired with the Atacama Pathfinder Experiment (APEX). APEX is a
  collaboration between the Max-Planck-Institut fur Radioastronomie,
  the European Southern Observatory, and the Onsala Space
  Observatory.}). Targets were observed using on--the--fly scans with
$\le{}10\arcsec$ spacing between rows. The half--power beam width is
$22\farcs{}3$ at the observed frequency, but smoothing during the
on--the--fly data reduction increases the beam size to
$23\farcs{}7$. Two or more perpendicular coverages were obtained to
mitigate scanning stripes. Data were taken in 2014 on April~29, May~2,
and May~7 under good weather conditions with system temperatures
$\le{}180~\rm{}K$. Data from the FLASH$^+$ receiver were recorded with
the XFFTS backend at a velocity resolution of
$0.04~\rm{}km\,s^{-1}$. This setup produced data in the frequency
ranges $\nu/{\rm{}GHz}=[280.0,282.5]$, $[292.0,294.5]$,
$[450.1,452.6]$, and $[463.6,466.1]$, but here we only exploit the
information on the $\rm{}N_2H^+$ line. Data reduction includes
standard calibration procedures and the subtraction of baselines of
second order. The spectra are gridded to a final velocity resolution
of $1.0~\rm{}km\,s^{-1}$.\medskip

\paragraph{Line Properties} Throughout this paper we adopt a rest
frequency of $279511.8~\rm{}MHz$ for the $\rm{}N_2H^+$ (3--2) line
\citep{caselli2002:l1544_i}. The hyperfine structure of the
$\rm{}N_2H^+$ (3--2) line is unlikely to have a significant impact on
the observed line structure. A compact summary is, for example,
provided by Table~A1 of \citet{caselli2002:l1544_i}. This demonstrates
that hyperfine satellites with velocity offsets exceeding
$\pm{}0.6~\rm{}km\,s^{-1}$ only contain a few percent of the
integrated relative intensities. In summary, satellites within
$\pm{}0.6~\rm{}km\,s^{-1}$ offset from the line reference frequency
might broaden the observed lines by about $1~\rm{}km\,s^{-1}$, but the
hyperfine structure is unlikely to have further impact on the observed
spectra.

The $\rm{}N_2H^+$ (3--2) transition has an upper--state energy of
$E_{\rm{}u}/k_{\rm{}B}=26.8~\rm{}K$, and an Einstein--$A_{ij}$
coefficient of $1.26\times{}10^{-3}~\rm{}s^{-1}$
\citep{mueller2001:cdms, muller2005:cdms}. This yields critical
densities of order $10^6~\rm{}cm^{-3}$ under typical CMZ conditions
with gas temperatures $\lesssim{}100~\rm{}K$
\citep{shirley2015:n_cr}. $\rm{}N_2H^+$ is the main workhorse tracer
of dense gas in molecular clouds with typical abundances
$\sim{}10^{-10}$ (e.g., \citealt{tafalla2006:internal-cores-ii}). The
molecule is believed to be primarily a tracer of gas that is cold
enough for depletion of CO on dust since $\rm{}N_2H^+$ is presumably
destroyed by CO \citep{caselli2002:n2h+}.

\subsection{Joint Imaging of Data\label{sec:joint-imaging}}
The interferometer and single--dish data are combined to produce maps
that contain information on both small and large angular scales. We
use separate processes for the continuum and emission line
data. Very few resources provide a hands--on
introduction on the combination of interferometer and single--dish
data. We therefore
continue to collect and document procedures at
\begin{equation*}
\text{\url{http://tinyurl.com/zero-spacing}}\quad{}.
\end{equation*}
Other relevant discussions of this problem are provided by
\citet{zhang2000:l1157}, \citet{kurono2009:zero-spacing}, and the IRAM
technical report\footnote{accessible at
  \url{http://www.iram-institute.org}} 2008--2 by Rodriguez-Fernandez,
Pety \& Gueth.

We apply a ``primary beam correction'' to all interferometer maps
analyzed here, in order to remove spatial gain variations in the
images, unless noted otherwise. The maps shown here do not include
this correction, though, since this suppressed noise at map
boundaries.\medskip

\begin{figure*}
\includegraphics[width=\linewidth]{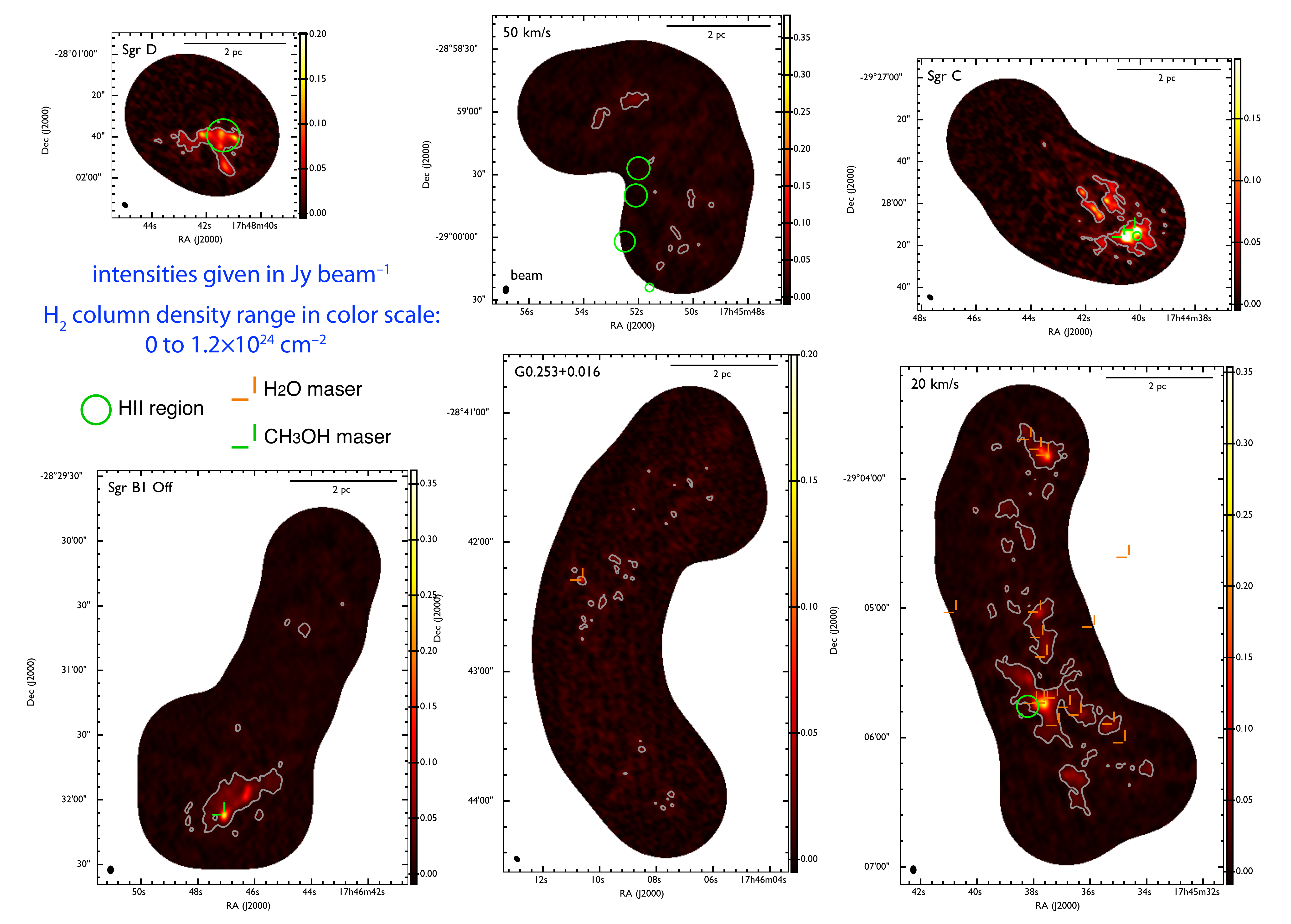}
\caption{Maps of dust continuum emission produced from the SMA
  interferometer data alone. Intensities are multiplied by gain
  distributions that suppress noise at the map boundaries. Beam shapes
  and sizes are indicated in the lower left corner of each panel,
  while a scale bar in the top right corner indicates physical
  dimensions. Color bars labeled in units of $\rm{}Jy\,beam^{-1}$ give
  the intensity scale separately for every image. Note that the beam
  size varies between maps. We choose the color bar to represent a
  fixed $\rm{}H_2$ column density range of
  $[0,1.2]\times{}10^{24}~\rm{}cm^{-2}$. In other words, images are
  represented in a manner that makes column densities directly
  comparable. A solid contour indicates the observed intensity exceeds
  the noise by a factor 5. Overlaid green circles, green crosses, and
  orange crosses indicate the locations of \ion{H}{ii}~regions, Methanol
  masers, and Water masers, respectively.\label{fig:maps-continuum}}
\end{figure*}

\begin{table*}
  \caption{Summary of image properties. The noise level for spectral
    line observations is determined for channels of
    $1~\rm{}km\,s^{-1}$ width. The bandwidth for continuum
    observations depends on the observing season, as indicated in
    Table~\ref{tab:observations-sma}. The position angle of the beam
    is measured East of North.\label{tab:image-properties}}
\begin{center}
\begin{tabular}{llllllllllllllllll}
\hline \hline
Target & \multicolumn{2}{c}{Setup} & Noise & Beam Size &
  Beam Position Angle\\ \hline
Sgr~C & 4~GHz & cont.\ & $6.5~\rm{}mJy\,beam^{-1}$ &
  $2\farcs{}6\times{}1\farcs{}6$ & $48\degr$\\
 & & $\rm{}N_2H^+$ & 39~mK &
  $5\farcs{}0\times{}4\farcs{}7$ & $124\degr$\\
$20~\rm{}km\,s^{-1}$ & 8~GHz & cont.\ & $3.9~\rm{}mJy\,beam^{-1}$ &
  $3\farcs{}4\times{}2\farcs{}2$ & $2\degr$\\
 & & $\rm{}N_2H^+$ & 36~mK & 
  $6\farcs{}0\times{}5\farcs{}2$ & $162\degr$\\
$50~\rm{}km\,s^{-1}$ & 8~GHz & cont.\ & $4.6~\rm{}mJy\,beam^{-1}$ &
  $3\farcs{}4\times{}2\farcs{}4$ & $177\degr$\\
 & & $\rm{}N_2H^+$ & 58~mK & 
  $5\farcs{}6\times{}4\farcs{}8$ & $159\degr$\\
G0.253+0.016 & 4~GHz & cont.\ & $3.5~\rm{}mJy\,beam^{-1}$ &
  $2\farcs{}5\times{}1\farcs{}7$ & $48\degr$\\
 & & $\rm{}N_2H^+$ & 42~mK & 
  $6\farcs{}7\times{}5\farcs{}9$ & $116\degr$\\
Sgr~B1--off & 8~GHz & cont.\ & $6.8~\rm{}mJy\,beam^{-1}$ &
  $3\farcs{}3\times{}2\farcs{}3$ & $0\degr$\\
 & & $\rm{}N_2H^+$ & 63~mK & 
  $5\farcs{}4\times{}4\farcs{}8$ & $163\degr$\\
Sgr~D & 4~GHz & cont.\ & $5.5~\rm{}mJy\,beam^{-1}$ &
  $2\farcs{}6\times{}1\farcs{}6$ & $50\degr$\\
 & & $\rm{}N_2H^+$ & 75~mK & 
  $3\farcs{}7\times{}3\farcs{}6$ & $42\degr$\\
\hline
\end{tabular}
\end{center}
\end{table*}

\noindent{}We use the Common Astronomy Software Applications package
(CASA\footnote{\url{http://casa.nrao.edu/}}) to image the dust
continuum emission. We build on the procedures of
\citet{zhang2000:l1157}.  Details of the processing are provided in
Appendix~\ref{sec-app:imaging-dust}. First, the interferometer
observations are imaged and cleaned. We improve the performance by
providing an iteratively improved source model to CASA's CLEAN
procedure via the ``modelimage'' option. We also use the
``multiscale'' option to suppress artifacts from mildly extended
emission. Second, the interferometer images are combined with
single--dish estimates of dust continuum. The latter data are
generated by scaling observations from the APEX Telescope Large Area
Survey of the Galaxy (ATLASGAL, \citealt{schuller2009:atlasgal}; see
\citealt{contreras2013:atlasgal-catalogue} for data reduction details)
that were done at a wavelength of $870~\rm{}\mu{}m$ using the Large
APEX Bolometer Camera (LABOCA;
\citealt{siringo2007:laboca,siringo2009:laboca}) to the observing
frequency of the SMA. See Appendix~\ref{sec-app:imaging-dust} for
details. We choose ATLASGAL data over the data provided by Bolocam
\citep{aguirre2011:bgps, bally2010:bolocam-gc} because of the higher
angular resolution. The FEATHER algorithm from CASA is used to execute
this combination. We stress that the data combination via FEATHER is
an approximate procedure that assumes perfect sampling of the
$uv$--plane out to the largest baselines observed by the
interferometers. Where possible we therefore rely on the
interferometer--only imaging products.

Figure~\ref{fig:maps-continuum} presents the maps of dust continuum
emission for this study. Indicators of star formation are overlaid
following the information compiled in
Appendix~\ref{sec-app:sf-observations}. Noise levels and beam shapes
are summarized in Table~\ref{tab:image-properties}. Note that
Fig.~\ref{fig:maps-continuum} shows maps without single--dish data
folded in. The data are also not corrected for the aforementioned gain
distributions. This suppresses signal but also noise at the map
boundaries. We choose this representation in order to reflect the
quality of the interferometer data. The intensities in
Fig.~\ref{fig:maps-continuum} are scaled to highlight a fixed column
density range for all clouds. The conversion between dust intensities
and column densities follows Appendix~A from
\citet{kauffmann2008:mambo-spitzer}. We adopt dust temperatures of
$20~\rm{}K$ (see Sec.~\ref{sec:obs-herschel}) and
\citet{ossenkopf1994:opacities} dust opacities for thin ice mantles
that have coagulated at a density of $10^6~\rm{}cm^{-3}$ for
$10^5~\rm{}yr$, as approximated by
\citet{battersby2011:cluster-precursors}, that are \emph{reduced by a
  factor 1.5} to be consistent with our previous work (see
\citealt{kauffmann2010:mass-size-i}).

The maps are largely devoid of significant emission. This is a main
feature of G0.253+0.016 already reported by
\citet{kauffmann2013:g0.253}. The new observations now show that this
relative absence of bright continuum emission on small spatial scales
is a general feature of CMZ molecular clouds. We defer a detailed
discussion of this trend to \citetalias{kauffmann2016:gcms_ii} that is focused on the cloud
density structure. In that publication we demonstrate that
$\lesssim{}10\%$ of the total cloud mass typically resides in
structures detected by the SMA.\medskip

\noindent{}We jointly image interferometer and single--dish
$\rm{}N_2H^+$ line emission data to recover both compact and extended
emission. To do this we start by combining the data in the
$uv$--domain. This requires a complex pipeline that is implemented
using MIRIAD and further explained in
Appendix~\ref{sec-app:imaging-lines}. In brief, we process the
single--dish maps to generate artificial data points in the
$uv$--plane that represent emission on large spatial scales. We then
combine these data with the regular $uv$--plane data points from the
interferometer. Weighting functions are used to produce a blended data
set that represents emission on both small and large scales. These
combined data are then imaged and cleaned to produce a final
image. This detailed combination of the single--dish and
interferometer data produces a larger restoring beam than that
obtained for interferometer--only data (i.e., in dust continuum).

This procedure produces good maps of all GCMS targets, except for
selected regions and velocity channels in the $20~\rm{}km\,s^{-1}$
cloud. Specifically we obtain maps containing strong corrugations in
some northern parts of this cloud. We attribute this to bright
$\rm{}N_2H^+$ emission outside of the region mapped by the SMA. In
this case too much information is missing to properly interpret the
signal picked up by the interferometer. The properties quoted for the
$20~\rm{}km\,s^{-1}$ cloud might be biased due to these problems, in
particular if measurements are made for compact objects. But we do not
think that these problems influence the conclusions obtained in this
paper. Note that the dust continuum emission is well covered by our
maps and consequently these data do not suffer from imaging
artifacts.

\subsection{Dust Observations with Herschel\label{sec:obs-herschel}}
We use data from the Herschel Space Telescope to provide a context on
cloud density structure within which the interferometer data obtained
here can be interpreted. Dust emission images of the region
$|\ell|\lesssim{}2\degr$ are used for this purpose. We also use
Herschel observations to obtain similar data on the cloud density structure of
the Orion~A molecular. This information is used to obtain reference
data against which the structure of CMZ clouds can be explored. The
Orion data are primarily exploited in \citetalias{kauffmann2016:gcms_ii}. The processing of
these maps is described here to provide a clear picture of Herschel
data reduction principles.

We use public Herschel data observed as part of the Herschel Infrared
Galactic Plane Survey \citep[Hi--GAL,][]{molinari2010:higal} and the
Gould Belt Survey \citep[GBS,][]{andre2010:herschel-gb} key
projects. Hi--GAL and the GBS cover, among other targets, the CMZ and
the Orion molecular cloud, respectively.
Table~\ref{tab:herschel-observations} indicates the target name, the
observation IDs, observing dates, and the associated projects of the
data used in this work.  All these observations were taken using the
parallel mode in which five wavebands (70, 160, 250, 350, and
$500~\rm{}\mu{}m$) are observed simultaneously. For the CMZ field we
also include maps from the ATLASGAL project on APEX
\citep{schuller2009:atlasgal, contreras2013:atlasgal-catalogue}. These
long--wavelength data cannot be included in the Orion analysis due to
strong spatial filtering of the bolometer observations.

\begin{figure}
\includegraphics[width=\linewidth]{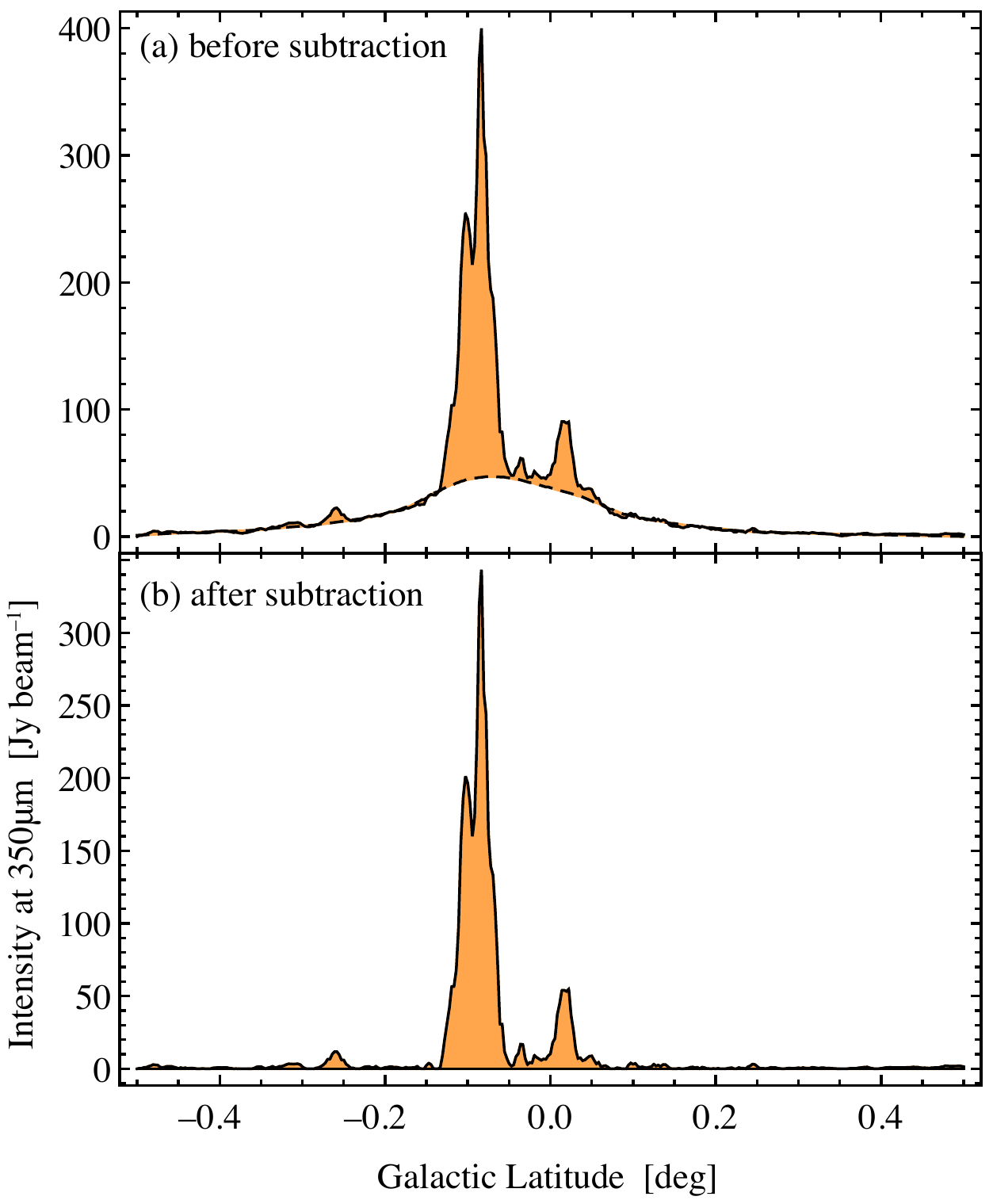}
\caption{Example of the background subtraction at
  $\ell=359\fdg{}87$. The most conspicuous feature corresponds to the
  $20~\rm{}km\,s^{-1}$ cloud.\label{fig:herschel-background}}
\end{figure}

Herschel data for the CMZ are reduced and processed following the
procedure described in \citet{guzman2015:herschel-malt90}.  Mapping
and destriping are done using standard tools provided by the Herschel
Interactive Processing Environment (HIPE) version 10. Maps are
convolved to the resolution of the $500~\rm{}\mu{}m$ maps (i.e.,
$37\arcsec$) by using ad--hoc convolution kernels
\citep{aniano2012:dust-starlight} and transformed to a common pixel
grid.  The procedure described in \citet{guzman2015:herschel-malt90}
includes the subtraction of a smooth component representing diffuse
emission from the Galactic Plane.  We refer to this diffuse component
as a ``background'', bearing in mind that it could equally represent
background or foreground emission.  The objective of this background
subtraction is twofold: (\textit{i})~to match the subtraction of the
low spatial frequency signal that affects the ATLASGAL data, and
(\textit{ii})~to subtract diffuse emission that cannot be attributed
to the individual CMZ clouds that are the targets of this study. As an
example, we show the performance of the background subtraction around
the $20~\rm{}km\,s^{-1}$ molecular cloud. This cloud is located very
near the Galactic Center, where significant diffuse emission is
present.  Figure~\ref{fig:herschel-background} shows the
$350~\rm{}\mu{}m$ intensity measured by Herschel versus the Galactic
latitude across the approximate longitude of the $20~\rm{}km\,s^{-1}$
molecular cloud ($\ell=359\fdg{}87$). It is apparent that the
subtracted background is not affecting the molecular cloud emission,
and conversely, not subtracting this component could induce one to
overestimate the column density of the CMZ clouds (or of any molecular
cloud located in the Galactic plane).  Moreover, the background
subtraction also corrects for the unknown offset of the Herschel
photometry which is shifted by an unknown quantity independent for
each field.

Background--subtracted images are combined and a single temperature
dust emission model is fitted to the intensities from 160 to
$870~\rm{}\mu{}m$ for each pixel.  The main difference between the
procedure employed in this work and the one described in
\citet{guzman2015:herschel-malt90} is that we base our assumed
description of the dust opacity prescription on the properties listed
by \citet{ossenkopf1994:opacities}. Here we assume thin ice coatings
and dust coagulation for $10^5~\rm{}yr$ at a molecular volume density
of $10^6~\rm{}cm^{-3}$, and we further decrease the resulting opacity
by a factor 1.5. This approach is chosen to be consistent with our
previous work on dust--based mass estimates (see
\citealt{kauffmann2010:mass-size-i}). We note that our
masses are slightly lower than those \citet{walker2015:dust-ridge}
quote for the same regions. The reason for this remains unclear, in
particular given that the lower dust opacities assumed by us should
yield higher masses for given flux.

Data for the Orion~A molecular cloud are reduced in a similar
way. These data are not exploited in the current publication, but they
play a major role as a reference data set in other GCMS publications.
Note that we confine the analysis to wavelengths between 250 and
$500~\rm{}\mu{}m$. We do so because visual inspection suggests that
some of the emission at lower wavelengths is not associated with dense
gas. For example, due to excitation from the OB--type stars in the
Orion Nebula, a significant fraction of the intensity might be due to
emission from very small grains that are not tracing the dense
interior of the cloud \citep{schnee2008:dust-perseus}. The calibrated
data sets are obtained from the Herschel Science Archive (Standard
Product Generation version 9.2.0). The mapping, destriping, and
combination of the three fields is carried out using Unimap
\citep{piazzo2015:unimap}. As done for the CMZ, we also subtract a
``background'' in order to distinguish Galactic plane diffuse emission
from the molecular cloud (and simultaneously correct for the
photometric offset).  For Orion~A, however, we consider a much
smoother background compared to those subtracted to the CMZ
fields. While the CMZ background filters out uniform emission on
scales of $\sim{}2\farcm{}5$, the Orion~A background subtraction
filters out uniform emission on scales of $\sim15\arcmin$.  The Orion
A far--IR Herschel maps are convolved, re--gridded, and column density
and temperature maps are fitted to the intensities in a similar
fashion as done with the CMZ.  The Herschel data for Orion have
previously been imaged by, e.g.,
\citet{lombardi2014:herschel-planck-orion} using alternative data
processing algorithms.

We use the Herschel data to obtain two sets of mass measurements for
every CMZ cloud. First, we characterize a large column density contour
that still clearly selects the target cloud. In practice we extract
all clouds at an $\rm{}H_2$ column density of
$9\times{}10^{22}~\rm{}cm^{-2}$ to do this. For this contour we obtain
the enclosed mass and the effective radius,
$r_{\rm{}eff}=(A/\pi)^{1/2}$, from the enclosed area, $A$. Second,
within this contour we determine for every target cloud the maximum
mass contained in a Herschel beam of $37\arcsec$, corresponding to
$r_{\rm{}eff}=0.74~\rm{}pc$. Results are collected in
Table~\ref{tab:target-summary}.

We obtain dust temperatures in the range 17 to 25~K for our target
clouds. This is consistent with previously reported Herschel--based
temperatures \citep{longmore2011:m025}. Assuming a spherical geometry,
uniform density, and a mean molecular weight per $\rm{}H_2$ molecule
of 2.8 proton masses \citep{kauffmann2008:mambo-spitzer}, we evaluate
the mean cloud densities as
$n({\rm{}H_2})=3.5\times{}10^4~{\rm{}cm^{-3}}\cdot
(M/10^4\,M_{\sun})\cdot(r/\rm{}pc)^{-3}$. The cloud masses and radii
reported in Table~\ref{tab:target-summary} give average densities of
$(0.9~{\rm{}to}~1.8)\times{}10^4~\rm{}cm^{-3}$ for the target clouds.


\begin{table}
\caption{Summary of the Herschel dust continuum observations used in
  this study.}
\label{tab:herschel-observations}
\begin{tabular}{cccc}
\hline \hline
Target Name  & Observation IDs & Obs.\ Date & Project\\\hline
Field-000\_0 & 1342204102/3 &  2010--09--07  & Hi--GAL\\
Field-002\_0 & 1342204104/5 &  2010--09--07  & Hi--GAL \\
Field-004\_0 & 1342214761/2 &  2011--02--24 & Hi--GAL\\
OrionA--N--1  & 1342218967/8 &  2011--04--09   & GBS\\
OrionA--C--1  & 1342204098/9 &  2010--09--06 & GBS\\
OrionA--S--1  & 1342205076/7 &  2010--09--26 & GBS\\
\hline
\end{tabular}
\end{table}

\section{Narrow Lines \& Linewidth--Size
  Relation\label{sec:kinematics}}
A major surprise of the \citet{kauffmann2013:g0.253} study of
G0.253+0.016 was the detection of very narrow lines with widths
$\le{}1~\rm{}km\,s^{-1}$ in structures with spatial scales
$\lesssim{}0.1~\rm{}pc$. This is remarkable given the strong unordered
gas motions of $\ge{}10~\rm{}km\,s^{-1}$ on the largest spatial scales
(e.g., \citealt{lis1994:m0.25}).

Here we take a thorough look at the single--dish and interferometer
data to develop a comprehensive view of CMZ cloud kinematics. This
work employs the well--detected $\rm{}N_2H^+$ (3--2) transition. Here
we use the combined data of APEX and the SMA for our analysis. We find
that the results of \citet{kauffmann2013:g0.253} are not affected by
spatial filtering and are representative of the entire CMZ.

\subsection{Properties of \sansmath{}$N_2H^+$ Maps\label{sec:n2h+-properties}}
\begin{figure}
\includegraphics[width=\linewidth]{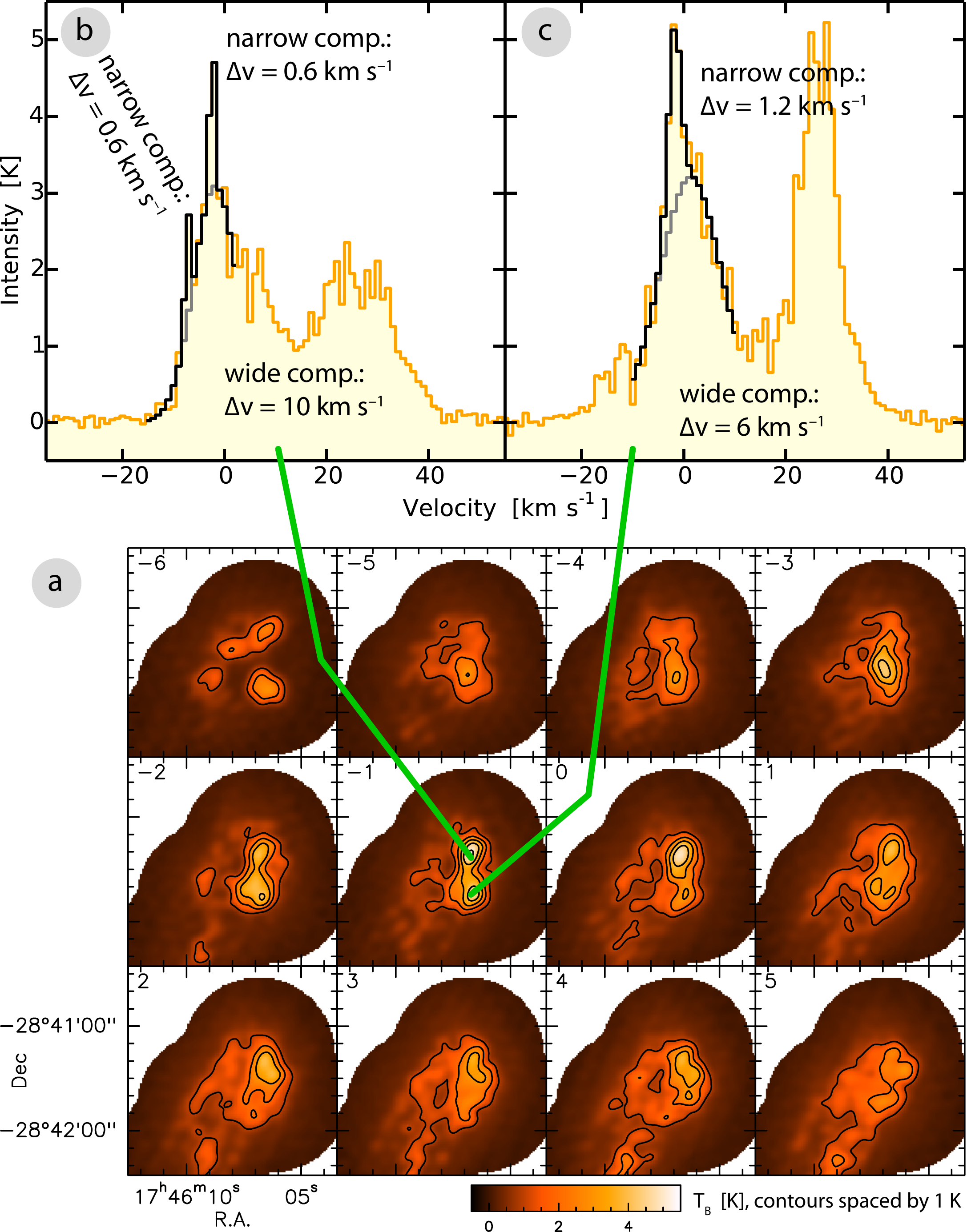}
\caption{$\rm{}N_2H^+$ (3--2) data for a sample region in the
  north of G0.253+0.016. \emph{Panel~a} presents channel maps. The
  velocity in $\rm{}km\,s^{-1}$ is indicated in the upper left
  corner. \emph{Panels~b and c} show spectra extracted towards
  selected positions. \emph{Black lines} in these plots depict
  multi--component Gaussian fits to selected velocity ranges of the
  spectra, where \emph{gray lines} trace the structure of the widest
  component. For each component we indicate the full width at half
  maximum, $\Delta{}v$.\label{fig:channels-g0253}}
\end{figure}

\begin{figure}
\includegraphics[width=\linewidth]{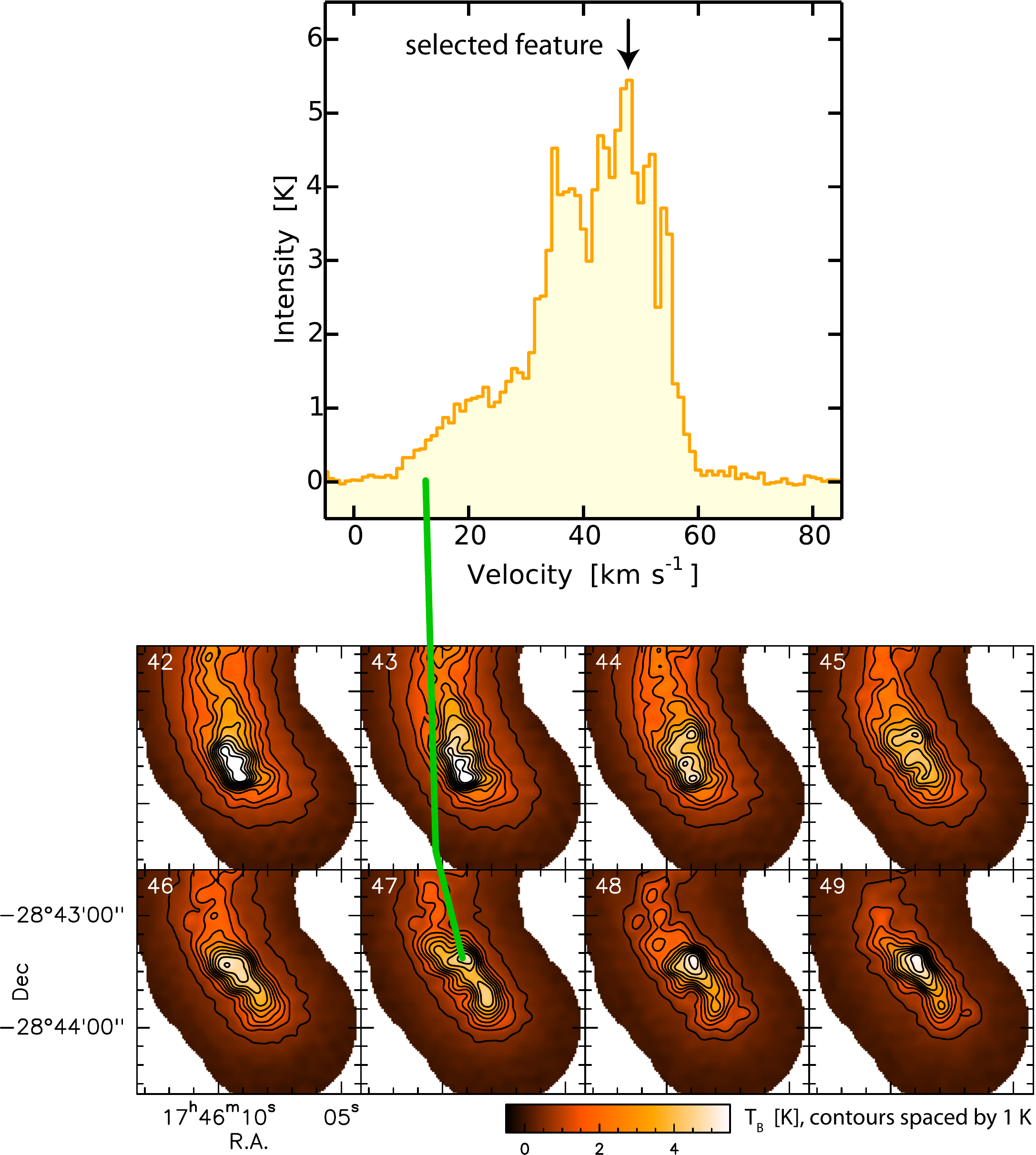}
\caption{As Fig.~\ref{fig:channels-g0253}, but for a southern section
  of G0.253+0.016. Map and spectrum indicate a feature selected in the
  dendrogram--based search for compact features in the $\rm{}N_2H^+$
  data.\label{fig:channels-g0253-south}}
\end{figure}

The $\rm{}N_2H^+$ maps are very rich in kinematic
structure. Figures~\ref{fig:channels-g0253} and
\ref{fig:channels-g0253-south} give an impression of the complexity
found in the maps, while Table~\ref{tab:image-properties} summarizes
aspects of the image quality.

Panel~(a) of Fig.~\ref{fig:channels-g0253} zooms in onto the northern
end of G0.253+0.016. Note the significant variations between velocity
channels. This is remarkable, given that the channels have a width of
only $1~\rm{}km\,s^{-1}$. Panels~(b) and (c) present spectra for two
of the intensity peaks seen in the map. In combination these panels
reveal that features of very narrow line width --- sometimes
comparable to the width of individual velocity channels --- are
superimposed on structures with much higher line width. This is
explicitly demonstrated via simultaneous Gaussian fits to parts of the
spectra in Fig.~\ref{fig:channels-g0253}. Specifically, the spectra
reveal two features with line widths that are too small to be properly
resolved by our channels of $1~\rm{}km\,s^{-1}$ spacing. But they also
expose several additional velocity components, including features with
widths of $6~{\rm{}to}~10~\rm{}km\,s^{-1}$ on which the more narrow
lines are superimposed.

Figure~\ref{fig:channels-g0253-south}, presenting data towards the
southern part of G0.253+0.016, reveals a qualitatively different
structure in the kinematic information. Here we find a much
more complex spectrum with multiple velocity components of relatively
narrow width crowding within a velocity range of about
$20~{\rm{}to}~30~\rm{}km\,s^{-1}$ width. Also note the difference in
bulk velocity compared to Fig.~\ref{fig:channels-g0253}.

Data of the sort depicted in Figs.~\ref{fig:channels-g0253} and
\ref{fig:channels-g0253-south} reveal three broad trends in cloud
structure: (\textit{i})~the existence of very narrow components that
are barely resolved by our channels with $1~\rm{}km\,s^{-1}$ spacing;
(\textit{ii})~the presence of broader velocity components with a width
$\lesssim{}10~\rm{}km\,s^{-1}$ (which might well be superpositions of
several unresolved components)); and (\textit{iii})~the fact that
several well--separated relatively wide velocity components can exist
along the same line--of--sight. The latter trend is for example
evident in the northern part of G0.253+0.016
(Fig.~\ref{fig:channels-g0253}[c]), where two broad features separated
by $\sim{}30~\rm{}km\,s^{-1}$ with peak intensities $\sim{}5~\rm{}K$
are evident.

The images also reveal a low level of imaging artifacts in the form of
``corrugations''. The maximum amplitude of these artifacts is
estimated manually by inspecting the depth of the minima between
consecutive peaks that appear to follow the corrugation pattern. This
yields a very conservative upper limit of $\le{}1~\rm{}K$ for
the imaging artifacts. As mentioned before, these artifacts are larger
for the northern part of the $20~\rm{}km\,s^{-1}$ cloud where bright
emission outside the map prevents a clean image reconstruction
(Sec.~\ref{sec:joint-imaging}).\medskip

\noindent{}The comparison of the $\rm{}N_2H^+$ and dust emission maps
reveals a surprising trend: there is no close correspondence between
the two on the angular scales probed by the interferometer. Some of
the dust emission peaks seen at high angular resolution are also seen
as peaks in $\rm{}N_2H^+$ channel maps, but sometimes this requires
close inspection of the data, and bright $\rm{}N_2H^+$ features
usually have no obvious counterparts in dust emission. Better
agreement is found on larger angular scales probed by single--dish
data (e.g., \citealt{rathborne2014:g0253-malt90}).

Similar trends have, for example, been reported by
\citet{rathborne2015:g0253-alma}. That study focusses on lines in the
3mm band. Many of those emission lines are optically thick, and
\citet{rathborne2015:g0253-alma} argue that radiative transfer effects
could in part explain the lack of correlation between the line and
dust data. However, such an explanation seems unlikely for the
$\rm{}N_2H^+$ (3--2) lines explored here, which have higher excitation
(i.e., temperature and density) requirements and should have a much
lower optical depth. This leaves complex spatial variations in the
chemical abundance as another explanation for the disconnect between
maps of dust and line emission, as also suggested by
\citet{rathborne2015:g0253-alma}. Further, recall that the CMZ gas
temperatures of 50 to 100~K are distinct from
the dust temperatures $\sim{}20~\rm{}K$. These temperature differences
between gas and dust can also help to reduce the correlation between
molecular emission lines and dust emission.

\subsection{Segmentation of \sansmath{}$N_2H^+$
  Emission\label{sec:n2h+-segmentation}}
The complexity of the gas kinematics complicates the extraction of
cloud fragments. We therefore follow a mixed approach.

In one scheme we start by finding the maximum intensity for a given
cloud in position--position--velocity ($p$--$p$--$v$) space. We then
draw an iso--intensity surface in $p$--$p$--$v$ space at a threshold
intensity corresponding to $1/3$ of this peak intensity. Contiguous
regions within these surfaces are considered to be coherent
structures. We refer to these structures as ``clumps''. This search
and the object characterization are done using the
ASTRODENDRO\footnote{\url{http://www.dendrograms.org}} implementation
of the dendrogram segmentation algorithm described by
\citet{rosolowsky2008:dendrograms}. The \texttt{min\_value} parameter
is used to reject regions below $1/3$ of the peak intensity. Our
exploration below considers structures on all spatial scales and is
thus robust with respect to the exact choice for the threshold
intensity. The size of clumps is characterized via the effective
radius $r_{\rm{}eff}=(A/\pi)^{1/2}$, where $A$ is the area on the sky
that contains all volume elements belonging to an extracted
structure. Spectra integrated within the $p$--$p$--$v$ surfaces are
used to obtain bulk velocities and velocity dispersions. The latter
property is only calculated if the peak intensity of a clump exceeds
the threshold intensity for clump selection by a factor 2 (i.e.,
achieves $\ge{}2/3$ of the peak intensity in the map). This
restriction guarantees spectra with well--defined peaks and line wings
that can be characterized meaningfully by a velocity
dispersion. The detailed properties of extracted clumps are given
  in Table~\ref{tab:props-clumps}. We include measures of the
  spatial extent along the overall major and minor axes of the
  structures following \citeauthor{rosolowsky2008:dendrograms}
  (\citeyear{rosolowsky2008:dendrograms}; see their Eq.~2) and the
  position angle (PA) of the major axis measured East from North. The
number of extracted clumps is listed in
Table~\ref{tab:props-n2h+}. With the notable exception of
G0.253+0.016, all clouds have a rather simple structure that consists
of only one or two clumps.

\begin{figure}
\includegraphics[width=\linewidth]{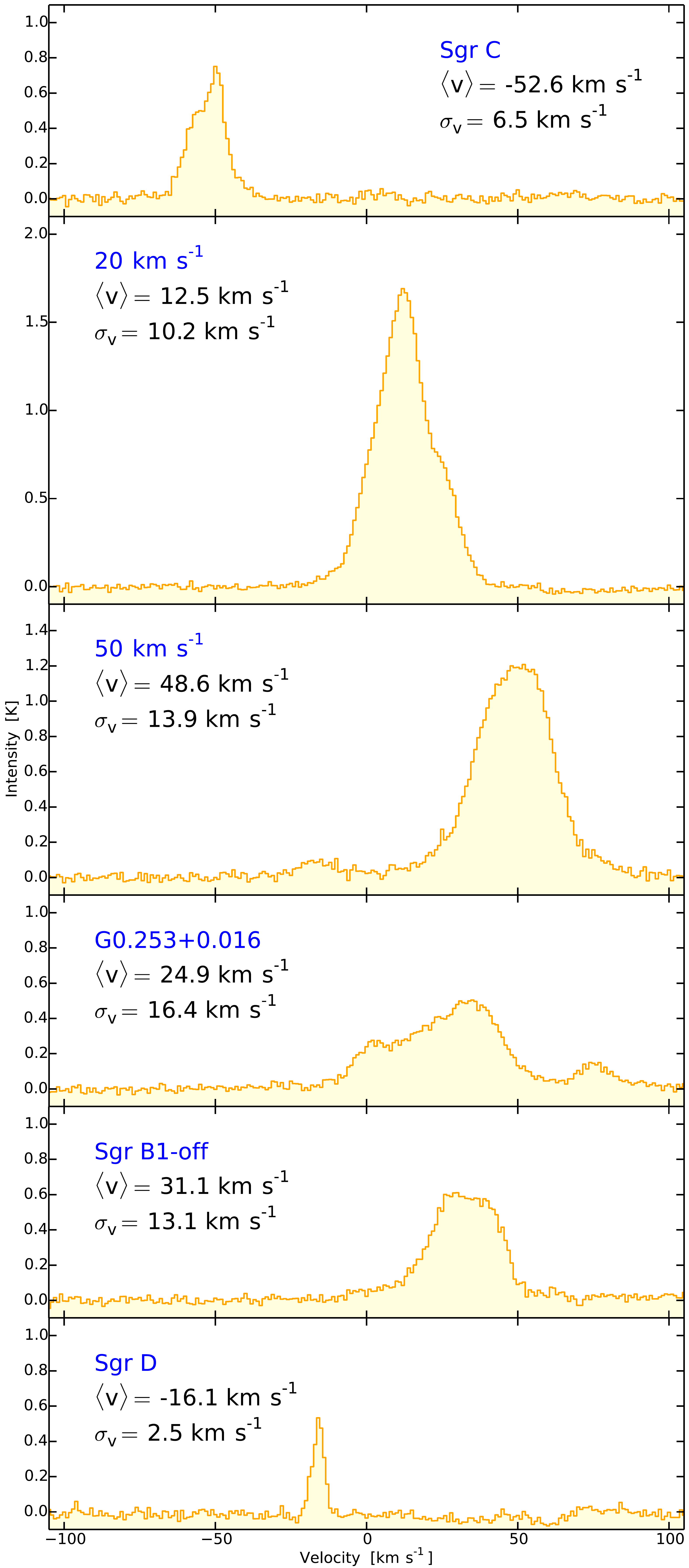}
\caption{Spatially averaged single--dish spectra of the $\rm{}N_2H^+$
  (3--2) transition for the target clouds. In each panel we indicate
  the mean velocity and velocity dispersion that are derived as the
  first and second moment of the spectra. See
  Secs.~\ref{sec:n2h+-segmentation} and \ref{sec:narrow-components}
  for details.\label{fig:integrated-spectra}}
\end{figure}

In a second scheme we use the single--dish data to obtain and
characterize integrated spectra for the clouds. These are shown in
Fig.~\ref{fig:integrated-spectra}. Table~\ref{tab:props-n2h+} lists
the mean velocities, $\langle{}v\rangle{}_{\rm{}SD}$, and velocity
dispersions, $\sigma_{\rm{}SD}(v)$, obtained as the first and second
moment of these spectra.

In a third scheme we focus on the most narrow lines in regions at the
resolution limit of our maps. This is described in the next section.

\onltab{
\begin{table*}
\begin{center}
\caption{Properties of dendrogram--extracted ``clumps'' as described
  in Sec.~\ref{sec:n2h+-segmentation}.\label{tab:props-clumps}}
\begin{tabular}{lcclcd{1}ld{1}llllllllllllllllllll}
\hline \hline
Region & RA & DEC & $r_{\rm{}eff}$ & size @ PA &
   \multicolumn{2}{l}{$\langle{}v\rangle{}$} &
   \multicolumn{2}{l}{$\sigma_v$}\\
 & \multicolumn{2}{c}{J2000.0} & pc & arcsec @ deg &
   \multicolumn{2}{l}{$\rm{}km\,s^{-1}$} &
   \multicolumn{2}{l}{$\rm{}km\,s^{-1}$}\\
\hline
Sgr~C & 17:44:41.1 & $-$29:28:09.3 & 0.7 & $9.1 \times 5.4$ @ 41 & -52.1 & & 3.0\\
$20~\rm{}km\,s^{-1}$ & 17:45:37.9 & $-$29:04:52.9 & 1.8 & $56.9 \times 10.3$ @ 14 & 13.6 & & 6.9\\
$50~\rm{}km\,s^{-1}$ & 17:45:51.0 & $-$28:59:21.8 & 1.2 & $24.5 \times 10.3$ @ 24 & 44.3 & & 6.4\\
G0.253+0.016 & 17:46:06.7 & $-$28:41:31.3 & 0.6 & $9.8 \times 4.5$ @ $-8$ & 0.6 & & 3.1\\
 & 17:46:07.2 & $-$28:41:26.9 & 0.8 & $8.5 \times 8.1$ @ 33 & 27.5 & & 4.1\\
 & 17:46:09.4 & $-$28:43:18.1 & 1.3 & $25.9 \times 8.8$ @ 12 & 40.8 & & 5.9\\
 & 17:46:09.7 & $-$28:42:32.4 & 0.5 & $6.8 \times 4.1$ @ $-35$ & 12.0 & & 2.6\\
 & 17:46:10.1 & $-$28:43:01.7 & 0.4 & $5.3 \times 3.4$ @ 9 & 20.9 & & 1.2\\
Sgr~B1 & 17:46:43.5 & $-$28:30:14.0 & 0.4 & $4.1 \times 3.8$ @ 57 & 34.1 & & 1.2\\
 & 17:46:46.8 & $-$28:31:57.0 & 1.0 & $11.2 \times 9.7$ @ $-46$ & 29.5 & & 4.4\\
Sgr~D & 17:48:41.0 & $-$28:01:41.2 & 0.1 & $1.7 \times 1.0$ @ 69 & -16.4 & & 0.6\\
 & 17:48:41.9 & $-$28:01:48.2 & 0.4 & $7.5 \times 4.3$ @ 60 & -14.9 & & 1.0\\
\hline
\end{tabular}
\end{center}
\end{table*}
}

\begin{table*}
  \caption{$\rm{}N_2H^+$ source
    properties. Subscripts ``SD'' refer to single--dish
    observations.\label{tab:props-n2h+}}
\begin{center}
\begin{tabular}{lll d{1} l d{1} l d{1} lllllllllll}
\hline\hline
Target & Significant Maxima & Clumps &
 \multicolumn{2}{l}{$\langle{}v\rangle{}_{\rm{}SD}$} &
 \multicolumn{2}{l}{$\sigma_{\rm{}SD}(v)$} &
 \multicolumn{2}{l}{$\min(\sigma_{\rm{}Int}[v])$}\\
 & & & \multicolumn{2}{l}{$\rm{}km\,s^{-1}$} &
 \multicolumn{2}{l}{$\rm{}km\,s^{-1}$} &
 \multicolumn{2}{l}{$\rm{}km\,s^{-1}$}\\
\hline
Sgr C & 7 & 1 & -52.5 & & 6.5 & & 1.1\\
$20~\rm{}km\,s^{-1}$ & 3 & 1 & 12.5 & & 10.2 & & 2.2\\
$50~\rm{}km\,s^{-1}$ & 9 & 1 & 48.6 & & 13.9 & & 1.1\\
G0.253+0.016 & 10 & 5 & 24.9 & & 16.4 & & 0.6\\
Sgr B1 off & 4 & 2 & 31.1 & & 13.1 & & 0.9\\
Sgr D & 4 & 2 & -16.1 & & 2.5 & & 1.5\\
\hline
\end{tabular}
\end{center}
\end{table*}

\subsection{Narrow Lines in CMZ Clouds\label{sec:narrow-components}}
We return to the quest for narrow lines in CMZ clouds, as initiated in
\citet{kauffmann2013:g0.253}. A particular issue is the systematic
characterization of the narrow lines, given the complex kinematics
described in Sec.~\ref{sec:n2h+-properties}. These narrow components
are in particular found towards prominent line emission peaks seen at
the highest angular resolutions.

In our search for such features we first find all significant local
maxima in $p$--$p$--$v$ space that are above the noise by a factor
$\ge{}5$. We also require that these maxima are separated from other
significant maxima by troughs with a minimum depth larger than the
amplitude of imaging artifacts (i.e., 1~K). In practice these criteria
are implemented via the \texttt{min\_value} and \texttt{min\_delta}
parameters in the ASTRODENDRO code. The extracted cloud fragments are
not well separated from the general cloud emission. For this reason we
do not attempt a characterization in terms of size and line width for
all these features. Instead we only report the number of significant
local maxima in Table~\ref{tab:props-n2h+}. We find a significant
variation between clouds. These numbers essentially characterize the
complexity of emission on relatively small $p$--$p$--$v$ scales.

We then use these data to obtain information on the narrowest line
widths prevailing in a given cloud. We manually search the spectra
towards the $p$--$p$--$v$ maxima to find for every cloud the most
narrow lines that are well--separated from other velocity
components. In a further step we characterize these spectra via
multi--component Gaussian fits of the sort illustrated in
Fig.~\ref{fig:channels-g0253}. The velocity dispersions found for the
narrowest components, $\min(\sigma_{\rm{}Int}[v])$, are listed in
Table~\ref{tab:props-n2h+}. Relatively narrow lines with velocity
dispersions in the range $0.6~{\rm{}to}~2.2~\rm{}km\,s^{-1}$ are found
in all CMZ clouds studied here. \citet{kauffmann2013:g0.253} and
\citet{rathborne2015:g0253-alma} initially reported such narrow lines
for G0.253+0.016. Now we can consider the prevalence of small line
widths on small spatial scales a general feature of all CMZ clouds.

The resulting velocity dispersions can be compared to the value
expected for one--dimensional thermal motions of a particle of mass
$m$ in gas at a temperature $T_{\rm{}gas}$,
\begin{equation}
\sigma_{{\rm{}th},m}(v) = 288 ~ {\rm{}m\,s^{-1}} \,
  \left( \frac{m}{m_{\rm{}H}} \right)^{-1/2} \,
  \left( \frac{T_{\rm{}gas}}{10~\rm{}K} \right)^{1/2}
\end{equation}
(where $m_{\rm{}H}$ is the hydrogen mass). For $\rm{}N_2H^+$ molecules
with $m=29\,m_{\rm{}H}$ at gas temperatures of 50 to 100~K, as
representative for the CMZ (see Sec.~\ref{sec:introduction}), one
obtains thermal velocity dispersions of
$0.11~{\rm{}to}~0.17~\rm{}km\,s^{-1}$. This means that non--thermal
``turbulent'' random gas motions dominate the observed velocity
dispersions. However, more relevant for the gas dynamics is the
comparison to the gas velocity dispersion for the mean free particle
with a mass $\langle{}m\rangle{}=2.33\,m_{\rm{}H}$ (see Appendix~A of
\citealt{kauffmann2008:mambo-spitzer}). This gives
$0.4~{\rm{}to}~0.6~\rm{}km\,s^{-1}$ for the 50 to 100~K temperature
range. This means that \emph{the non--thermal gas motions become
  similar to the thermal motions of the mean free particle}. In other
words, the one--dimensional sonic Mach number
\begin{equation}
\mathcal{M} = \sigma_{\rm{}obs}(v)/\sigma_{{\rm{}th},\langle{}m\rangle}(v)
\end{equation}
determined from the observed velocity dispersion along the line of
sight, $\sigma_{\rm{}obs}(v)$, indicates \emph{transonic to mildly
  supersonic} gas motions instead of highly supersonic (i.e.,
$\mathcal{M}\gg{}1$) ones. Specifically, observed velocity dispersions
of $0.6~{\rm{}to}~2.2~\rm{}km\,s^{-1}$ imply
$\mathcal{M}=1.5~{\rm{}to}~5.5$ at 50~K gas temperature. Values lower
by a factor $2^{1/2}\approx{}1.4$ apply for temperatures of 100~K. For
comparison, for 50~K gas temperature we obtain much larger sonic Mach
numbers in the range $\mathcal{M}=15~{\rm{}to}~40$ from single--dish
velocity dispersions $\sigma_{\rm{}SD}(v)$ as listed in
Table~\ref{tab:props-n2h+}.

\subsection{Steep Linewidth--Size Relation\label{sec:linewidth-size}}
The combination of interferometer maps and single--dish observations with
a multi--scale characterization of cloud kinematics permits us to
construct the linewidth--size relation for the CMZ. Here we construct
the first such relation that extends from spatial scales
$\sim{}0.1~\rm{}pc$ to radii $>1~\rm{}pc$ (see
\citealt{rathborne2015:g0253-alma} for a multi--tracer exploration
that characterizes scales $\ge{}0.5~\rm{}pc$ in a single cloud).

\begin{figure}
\includegraphics[width=\linewidth]{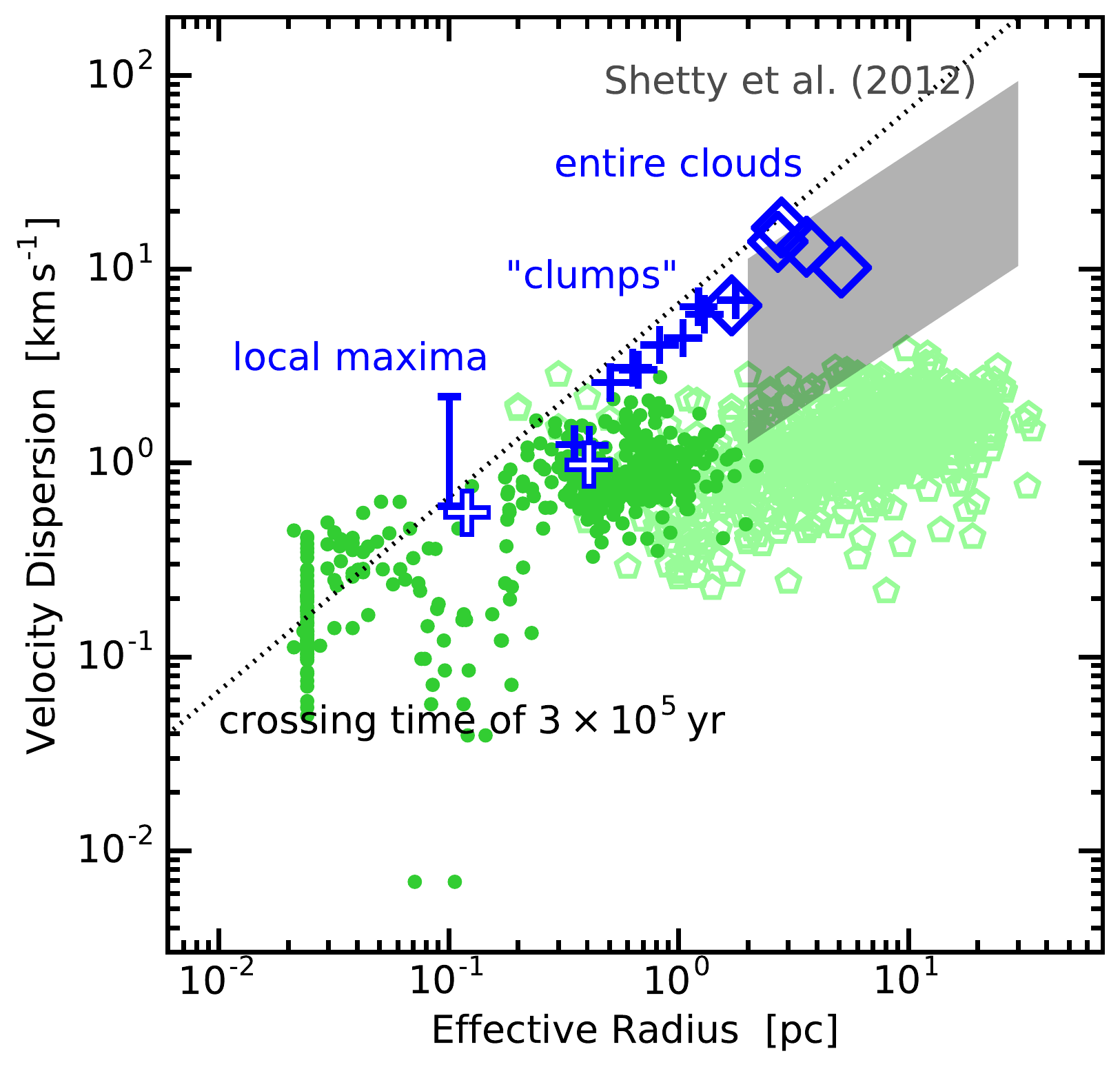}
\caption{Summary of the $\rm{}N_2H^+$--based linewidth--size data
  explored in this paper. \emph{Blue symbols} give characteristics of
  structures of varying spatial size explored here; see
  Sec.~\ref{sec:linewidth-size} for terminology and details. The
  \emph{blue crosses with white filling} indicate data for the Sgr~D
  region that probably resides outside the CMZ as studied here.  The
  \emph{gray shaded region} summarizes the linewidth--size
  measurements reported by \citet{shetty2012:linewidth-size-cmz}.
  \emph{Dark green bullets} indicate the properties of reference Milky
  Way clouds from the \citet{kauffmann2013:virial-parameter}
  compilation that is based on emission lines of $\rm{}N_2H^+$ and
  $\rm{}NH_3$ that trace dense gas. \emph{Light green diamonds} give
  CO--based data for the lower density gas in Milky Way clouds
  reported in the same collection. The \emph{dotted line is defined by
    a crossing time of
    $3\times{}10^5~\rm{}yr$.}\label{fig:linewidth-size}}
\end{figure}

Figure~\ref{fig:linewidth-size} collects the information on gas
kinematics obtained above. The velocity dispersions from the
single--dish spectra are combined with the cloud size measurements
from Herschel given in Table~\ref{tab:target-summary}. The ``clumps''
refer to the larger cloud structures identified in the dendrogram
analysis. The vertical bar indicates the range of minimum velocities
found in the clouds by fitting spectra towards individual local
$p$--$p$--$v$ maxima. A radius of $2\farcs{}5$ (i.e., 0.1~pc) is
adopted for these structures, roughly corresponding to the beam radius
of the observations. Figure~\ref{fig:linewidth-size} also includes
information on CMZ gas kinematics inferred by
\citeauthor{shetty2012:linewidth-size-cmz}
(\citeyear{shetty2012:linewidth-size-cmz}; i.e., a factor 3 around the
$\rm{}N_2H^+$ results in their Table~2) and the non--thermal
kinematics of Milky Way molecular clouds inferred from CO and dense
gas tracers compiled in \citet{kauffmann2013:virial-parameter}.

We note that the measurements of size and velocity dispersion used in
Fig.~\ref{fig:linewidth-size} suffer from negligible uncertainties due
to noise and imaging artifacts. We have tested this by
(\textit{i})~repeating the measurements in data sets to which we add
artificial noise at the observed level. The measurements change at the
level of a few percent between different initializations of the random
noise. We have also (\textit{ii})~examined results for
dendrogram--based extractions where the \texttt{min\_delta} parameter
was set to 1~K, i.e., the maximum level of the ``corrugations'' seen
in our images. These experiments show that the structure segmentation
is not influenced by the imaging
artifacts. Figure~\ref{fig:linewidth-size} does therefore not include
uncertainty bars.

We recover the well--known trend that CMZ molecular clouds have very
high velocity dispersions when explored on large spatial scales. For
example, CMZ gas velocity dispersions measured on about 5~pc scale
exceed the values for clouds elsewhere in the Milky Way by about an
order of magnitude. However, we also find that low velocity
dispersions on small spatial scales are prevalent in all clouds. It
appears that the observed CMZ gas motions are at the upper end of what
is usually observed for spatial scales $\sim{}0.1~\rm{}pc$ elsewhere
in the Milky Way, but the dense gas kinematics in the CMZ are not
entirely different from what is found in regular Milky Way clouds.

A very surprising trend is the steep linewidth--size relation seen in
the data. We fit the observed relation by a law
\begin{equation}
\sigma(v) = \sigma_0 \, (r_{\rm{}eff}/{\rm{}pc})^{h_{\sigma(v)}}
\end{equation}
in order to obtain a quantitative description. Specifically, we
minimize the mean squared residuals between the base 10 logarithm of
the observed and predicted velocity dispersions. Our analysis does not
yield good estimates for the systematic uncertainties of our
observations of velocity dispersion. Uncertainties in $h_{\sigma(v)}$
are therefore estimated as follows. We ignore linewidth--size data on
the ``clumps'' for this analysis, since these data anyway lie between
those for entire clouds and local maxima and thus add little
additional information. We then vary the observed velocity dispersion
on the smallest spatial scales in the range
$0.6~{\rm{}to}~2.2~\rm{}km\,s^{-1}$, i.e., between the extremes of the
observed values. We assume that these two fits bracket the true range
of linewidth--size relations. This gives
\begin{equation*}
\sigma(v) = (5.5\pm{}1.0)~{\rm{}km\,s^{-1}} \,
  (r_{\rm{}eff}/{\rm{}pc})^{0.66\pm{}0.18}
\end{equation*}
as the best fit. Interestingly, \citet{shetty2012:linewidth-size-cmz}
report a similar value of 0.67 for this slope, based on the
$\chi^2$--minimization of measurements from sigle--dish data, but
their confidence range spans from 0.41 to 1.13 and is much larger than
ours (arguably due to a more consequent treatment of uncertainties).

We find that \emph{the observed linewidth--size slope is unusually
  steep}, compared to other studies of Milky Way clouds. It is not
straightforward to establish this trend: one issue is that different
data sets are processed in different ways. A reasonable reference is
given by the collection of cloud kinematics in
\citet{kauffmann2013:virial-parameter} which focusses on tracers of
dense gas similar to what is done here. That study yields
$\sigma(v)\propto{}r_{\rm{}eff}^{0.32}$, which is much flatter than
the relation found here. \citet{kauffmann2013:virial-parameter} does
not list an uncertainty for the slope since that was not needed in
that work. Still, it seems to be plausible that the linewidth--size
relation for the CMZ is by a significant margin steeper than the
relation for the dense gas in other Milky Way clouds.

Another reference point for linewidth--size relations is the classical
$\sigma(v)\propto{}r_{\rm{}eff}^{1/2}$ relation for entire clouds
traced in CO, as inspired by the \citet{larson1981:linewidth_size}
study. The modern formulation, including the exponent, is derived in a
combination of observational and theoretical arguments (e.g.,
\citealt{heyer2015:review}). This leaves the uncertainties applying to
this relationship somewhat unclear. However, purely observational
determinations ranging from 0.4 to 0.6 (e.g.,
\citealt{sanders1985:gmcs_ii}, \citealt{solomon1987:co-survey},
\citealt{heyer2004:turbulence}) bracket this trend and suggest an
uncertainty of $\pm{}0.1$ for the slope. These findings suggest that
the CMZ linewidth-- size relation does not significantly deviate from
what is found for regular Milky Way clouds.

Still, it is apparent from Fig.~\ref{fig:linewidth-size} that the
velocity dispersions of CMZ clouds significantly exceed those of
regular Milky Way clouds for radii $\gtrsim{}1~\rm{}pc$, while no
clear difference is apparent at radii $\sim{}0.1~\rm{}pc$. We note
that this analysis combines data generated from tracers that are
potentially very different. While the small scales of regular Milky
Way clouds are characterized using tracers of dense gas similar to
those we use for the CMZ (i.e., $\rm{}N_2H^+$ and $\rm{}NH_3$; dark
blue bullets), the largest scales are probed using CO (light blue
diamonds), i.e., a molecule that traces gas at densities much below
those prevailing in the CMZ. It is, however, generally observed that
CO--based velocity dispersions significantly exceed those of
$\rm{}N_2H^+$ and $\rm{}NH_3$ (e.g.,
\citealt{kirk2010:dense-core-dynamics-ii}). We can thus confidently
claim that CMZ clouds explored at radii $\gtrsim{}1~\rm{}pc$ clearly
have higher velocity dispersions than regular Milky Way
clouds.\medskip

\noindent{}In summary the data in Fig.~\ref{fig:linewidth-size}
indicate that the linewidth--size relation of CMZ clouds is unusually
steep by a significant factor. We stress that \emph{our claims about
  an unusually steep linewidth--size relation in CMZ clouds chiefly
  rest on the manual inspection of the data points shown in
  Fig.~\ref{fig:linewidth-size}.}  We feel that this direct comparison
suffers from fewer uncertainties than the quantitative discussion of
linewidth--size fits to data sets. This unusually steep relation might
have important implications for the nature of turbulence in CMZ clouds
(Sec.~\ref{sec:dissipation}).

That said, we caution that \citet{shetty2012:linewidth-size-cmz} find
that the linewidth--size slope does not change between the CMZ and the
solar neighborhood. This is in particular remarkable because their
linewidth--size slope of 0.67 for $\chi^2$--minimized $\rm{}N_2H^+$
data is similar to our result. The
\citeauthor{shetty2012:linewidth-size-cmz} results are in agreement
with previous investigations by \citet{oka1998:co-cmz_ii}
\citet{miyazaki2000:dense-gas-cmz_ii} who also find that the
kinematics of CMZ molecular clouds deviate from those of regular Milky
Way clouds only by having a larger intercept, $\sigma_0$.  However,
our study differs from the \citeauthor{shetty2012:linewidth-size-cmz}
investigation in that we can compare clouds inside and outside the CMZ
\emph{at the same spatial scale}. We thus deem our results to be more
robust than the conclusions provided by
\citeauthor{shetty2012:linewidth-size-cmz}
\citeauthor{oka1998:co-cmz_ii} do not cover a large range of spatial
scales, and so they can only constrain
$\sigma_0$. \citeauthor{miyazaki2000:dense-gas-cmz_ii}, by contrast,
conduct an analysis very similar to ours, i.e., they compare CMZ
regions with regular Milky Way clouds over a range of spatial scales
in their Fig.~2(b). It is thus interesting that their analysis only
indicates differences in $\sigma_0$, while the slope appears to be
identical everywhere in the Milky Way. A major difference between
their and our work, though, is that we can access spatial scales
$\lesssim{}0.1~\rm{}pc$. The difference between our and their work
could thus be explained if the linewidth--size relation steepens on
small spatial scales.\medskip

\noindent{}The data in Fig.~\ref{fig:linewidth-size} also allow to
obtain the flow crossing time,
$t_{\rm{}cross}=\ell{}/\sigma(v)$. Undriven turbulence is supposed to
decay as ${\rm{}e}^{-t/t_{\rm{}cross}}$ \citep{maclow2004:review}, and
so $t_{\rm{}cross}$ also indicates the time scale on which turbulence
might decay. The actual decay time might exceed $t_{\rm{}cross}$,
however, if the medium has clumped up into cloud fragments with small
volume filling factor that interact little with another. We adopt
$\ell{}=2\,r_{\rm{}eff}$ to indicate where
$t_{\rm{}cross}=0.3~\rm{}Myr$ in Fig.~\ref{fig:linewidth-size}. The
velocity dispersions obtained for the local $\rm{}N_2H^+$ intensity
maxima are above this relation, and they give values as low as
$t_{\rm{}cross}=9\times{}10^4~\rm{}yr$. The kinematics for the larger
cloud structures are below the reference relation, implying
$t_{\rm{}cross}>0.3~\rm{}Myr$. These measurements set a reference time
scale, e.g., for our later discussion of star formation processes.

\section{Suppression of CMZ Star
  Formation\label{sec:suppression-sf}}
\begin{figure}
\includegraphics[width=\linewidth]{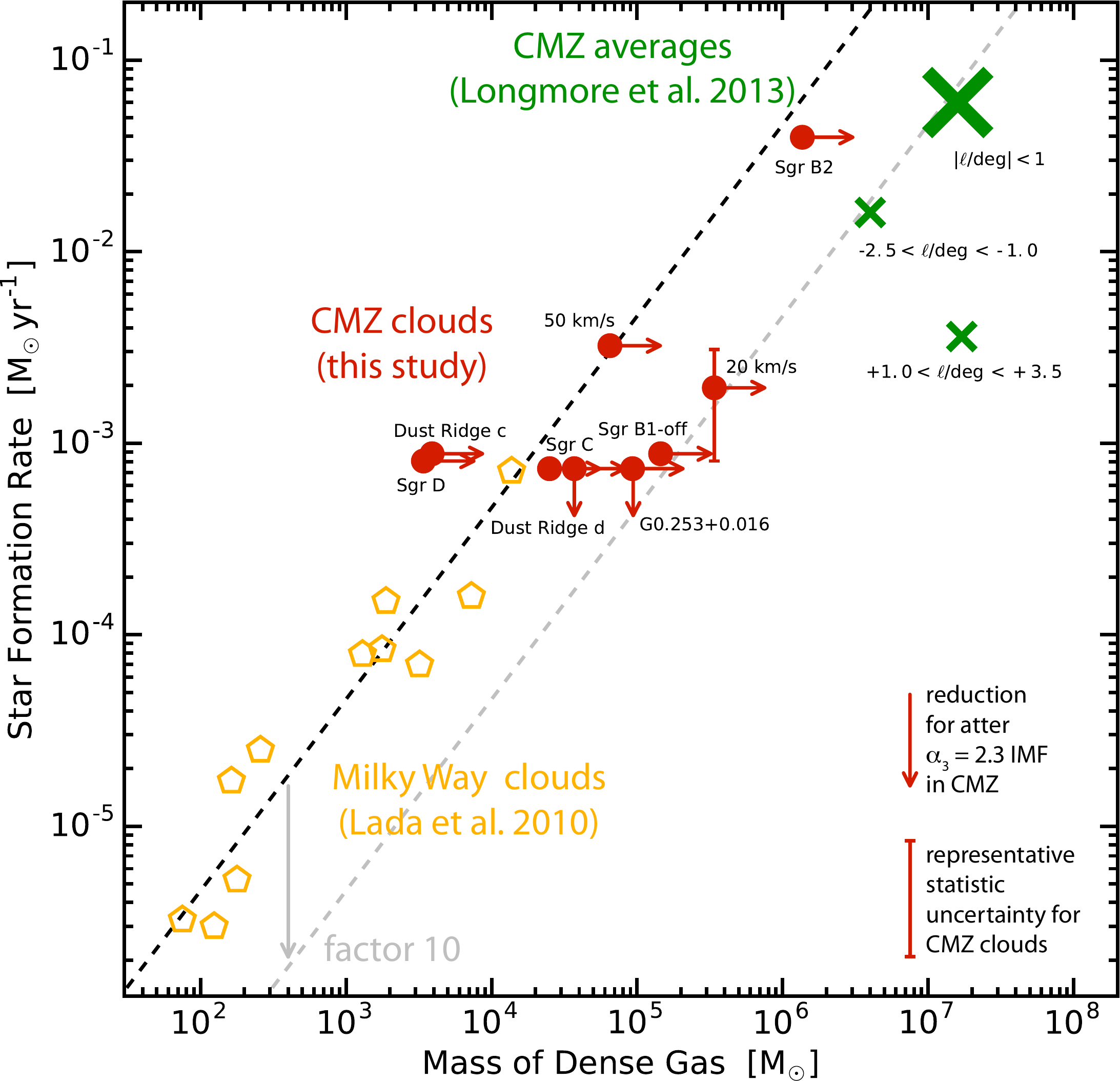}
\caption{Observed star formation rates vs.\ the mass of dense gas
  residing at visual extinctions $A_V>7~\rm{}mag$. \emph{Yellow
    symbols} give the properties of molecular clouds within about
  500~pc from sun compiled by
  \citet{lada2010:sf-efficiency}. \emph{Green crosses} plot data for
  several degree--sized regions in the CMZ from
  \citet{longmore2012:sfr-cmz}, as indicated by small labels. The
  \emph{large green cross} applies to the $|\ell{}|\le{}1\degr$ region
  largely explored by this paper. The \emph{red bullets} give masses
  and star formation rates for individual CMZ clouds determined in
  this paper building on data for methanol masers and \ion{H}{ii}
  regions. The \emph{red error bar} in the lower right corner
  indicates the statistical uncertainty in SF rate that holds for the
  CMZ clouds containing few \ion{H}{ii}~regions or none.  This
  excludes Sgr~B2, for which much lower statistical uncertainties
  apply. The nearby \emph{red arrow} indicates the reduction in SF
  rate for the same clouds with \ion{H}{ii}~regions if a flatter IMF
  is adopted. The \emph{black dashed line} indicates a fit to the
  solar neighborhood data taken from
  \citet{lada2010:sf-efficiency}. The \emph{gray dashed line} gives a
  relation with a star formation rate lower by a factor
  10.\label{fig:sf-mass}}
\end{figure}

As explained in Sec.~\ref{sec:introduction}, it has been known for
decades that star formation in the dense gas of the CMZ is suppressed
by an order of magnitude, compared to clouds elsewhere in the Milky
Way.  \citet{longmore2012:sfr-cmz} quantify this trend comprehensively
and provide a general framework in which the suppression of SF can be
discussed. Previous studies establish that the CMZ as a whole produces
fewer stars than one would expect based on the integral gas
properties. In \citet{kauffmann2013:g0.253} we demonstrate this
suppression quantitatively for a single \emph{individual}
cloud. Figure~\ref{fig:sf-mass} extends this analysis to include all
major molecular clouds in the CMZ\footnote{We remark that the
  discussion presented here constitutes a significant update to the
  analysis in \citet{kauffmann2013:g0.253}. First, we now build on own
  well--defined mass measurements for CMZ molecular clouds, while
  earlier we had to compile estimates from the literature. Second, we
  now consider the stellar initial mass function in more detail.}.

This discussion requires a joint analysis of information on both the
dense gas in clouds and the star formation activity. The more
technical details are presented in
Appendices~\ref{sec-app:sf-observations} and
\ref{sec-app:sf-rate}. Here we aim to summarize the discussion
presented there.

\subsection{Observations of Star Formation in CMZ Clouds
  \label{sec:sf-observations}}
Few systematic studies of star formation in CMZ clouds exist. For
example, to our best knowledge there is no reasonably complete and
reliable compilation that would catalogue \ion{H}{ii} regions as
reliable signposts of high--mass star formation. We therefore collect
this information from specific studies of selected target
clouds. Such observations are typically conducted with the VLA. We
build on work by \citet{liszt1992:sgr_survey} and
\citet{liszt1995:sgr_c} at 1.6~GHz ($\sim{}19~\rm{}cm$), and a series
of papers by \citet{mehringer1993:sgr_b1, mehringer1998:sgr_d} at
wavelength between 3 and 20~cm. Smaller fields were investigated by
\citet{ho1985:cmz_radio} at 2 and 6~cm, \citet{gaume1995:sgr_b2-i} at
1.3~cm, \citet{mills2011:continuum-50kms} at 8.4~GHz (3.6~cm),
\citet{immer2012:multi-wavelength-cmz} at 8.4~GHz (3.6~cm), and
\citet{rodriguez2013:g0.253} at 1.3 and 5.6~cm.  In addition, for the
Sgr~C region we rely on the focussed study by
\citet{kendrew2013:sgr-c}, who in part base their work on radio
continuum data from \citet{forster2000:radio-continuum}.

The radio continuum observations reveal a wealth of structure
including filaments that are not in any obvious way related to star
formation, supernova remnants (SNRs), as well as \ion{H}{ii} regions
of all kinds of sizes (i.e., from $>1~\rm{}pc$ to
$\ll{}1~\rm{}pc$). It is therefore prudent to inspect and discuss
regions individually. This is done in
Appendix~\ref{sec-app:sf-observations}.

The situation is better for class~II methanol ($\rm{}CH_3OH$)
masers. Here a systematic study exists in the form of the methanol
Multibeam Survey that employs the Parkes and ATCA observatories
\citep{caswell2010:gal-center-methanol-masers}. The data are less
complex than the radio continuum images and are thus easier to
interpret. Class~II methanol masers are believed to trace high--mass
star formation. However, since maser emission usually involves a
fortuitous viewing geometry, it is not clear what fraction of embedded
young high--mass stars is associated with these masers. In other
words, the presence of a class~II methanol maser is a sufficient
marker of an embedded high mass protostar, but not a necessary one.
Thus the use of class~II methanol masers as tracers of star formation
might be limited.

Water masers, e.g.\ near 22~GHz frequency, provide yet another probe
to detect embedded star formation. However, only the search by
\citet{lu2015:20kms} achieves a sensitivity of order 10~mJy that is
needed to reveal a significant population of the embedded
masers. Future deep searches with the VLA have the potential to change
this picture.

We summarize our results in Table~\ref{tab:sf-tracers}. We list only
the most relevant reference for each cloud and tracer. The methanol
masers are always taken from
\citet{caswell2010:gal-center-methanol-masers}. We consider clouds and
methanol masers to be associated only if the maser resides inside the
$9\times{}10^{22}~\rm{}cm^{-2}$ column density contour used to trace
cloud boundaries. Detailed descriptions of the data for every cloud
are provided in Appendix~\ref{sec-app:sf-observations}.

Figure~\ref{fig:maps-continuum} demonstrates that most masers are
clearly associated with compact and dense structures seen in dust
emission. This is expected since the masers should trace the youngest
stars that are still deeply embedded in their birth environments. The
\ion{H}{ii}~regions, by contrast, are typically offset from the denser
cloud material. This is consistent with the idea that the
\ion{H}{ii}~regions trace a more evolved stellar population.

\begin{table*}
\begin{center}
  \caption{Summary of the star formation activity in the
    clouds.\label{tab:sf-tracers}}
\begin{tabular}{lllllllllllll}
\hline \hline
Cloud & Methanol Masers & Radio Continuum \& Spectral Type &
Main Reference & $\dot{M}_{\rm{}SF}/(10^{-3}~M_{\sun}\,\rm{}yr^{-1})$\\ \hline
Sgr~C & 2 & one source & \citet{forster2000:radio-continuum} & 0.8\\
$20~\rm{}km\,s^{-1}$ & 0 & 1 $\times$ O9 &
\citet{lu2015:20kms} & 0.8--3.1\\
$50~\rm{}km\,s^{-1}$ & 0 & 1 $\times$ O7, 2 $\times$ O8.5, 1 $\times$
O9 & \citet{mills2011:continuum-50kms} & 3.2\\
G0.253+0.016 & 0 & none &
\citet{immer2012:multi-wavelength-cmz} & $<$0.8\\
Dust Ridge C & 1 & none & \citet{immer2012:multi-wavelength-cmz} &
$\sim{}$0.9\\
Dust Ridge D & 0 & none & \citet{immer2012:multi-wavelength-cmz} &
$<$0.8\\
Sgr~B1--off & 1 & none & \citet{mehringer1993:sgr_b1} & $\sim{}$0.9\\
Sgr~B2 & 11 & 20 $\times$ B0, 20 $\times$ O9.5 to O7, &
\citet{gaume1995:sgr_b2-i} & 36\\
 & & 9 $\times$ O6.5 to O5\\
Sgr~D & 0 & 1 $\times$ O7 & \citet{liszt1992:sgr_survey} & 0.8\\
\hline
\end{tabular}
\end{center}
\end{table*}

\subsection{Estimates of the Star Formation Rate\label{sec:sf-rate}}
The observations of star formation reviewed above only constrain the
presence of high--mass stars in the target clouds. But the presence
--- or absence --- of high--mass stars hints at the total stellar
population embedded in the clouds. This holds since the relative
numbers of stars of low and high mass are related via the initial mass
function (IMF). Appendix~\ref{sec-app:sf-rate} explains how these
observations can be combined with lifetime estimates of young stars to
calculate star formation rates. Here we provide a summary of that
discussion. Our discussion assumes a star formation rate of
$<0.06\,M_{\sun}\,\rm{}yr^{-1}$ for the region $|\ell{}|\le{}1\degr$
as derived by \citet{longmore2012:sfr-cmz} using free--free
foregrounds extracted from WMAP data.

In this study we adopt the \citet{kroupa2002:imf} IMF with
$\alpha_3=2.7$. Consider a stellar cluster that is described by this
IMF. Fitting of the \citeauthor{kroupa2002:imf} results with a power
law shows that the number of cluster members down to a mass of
$0.01\,M_{\sun}$, $N_{\rm{}cl}$, depends on the mass of the most
massive star in the cluster, $M_{\rm{}max}$, as
\begin{equation}
N_{\rm{}cl} \approx 20.5 \cdot (M_{\rm{}max} / M_{\sun})^{1.7} \, .
\label{eq:n_cl}
\end{equation}
We adopt the mean stellar mass
$\langle{}m_{\ast}\rangle{}=0.29\,M_{\sun}$ that holds for the
\citet{kroupa2002:imf} IMF with $\alpha_3=2.7$. Here we use these
relationships to estimate the total mass of the young stars embedded
in a cloud as
$\langle{}m_{\ast}\rangle{}\cdot{}N_{\rm{}cl}$. We thus
need to determine $N_{\rm{}cl}$.

\ion{H}{ii}~regions contain central stars of spectral types O9.5 or
earlier. These have stellar masses $\ge{}19\,M_{\sun}$ following
Table~4 of \citet{weidner2010:o-star-masses}. For a given cloud we
thus use Eq.~(\ref{eq:n_cl}) to predict the number of stars required
to form one \ion{H}{ii}~region. We then multiply the result with the
number of embedded \ion{H}{ii}~regions to estimate the population of
young stars associated with the cloud. Similarly, we assume
$M_{\rm{}max}\sim{}20\,M_{\sun}$ in case class~II methanol masers are
detected in clouds devoid of \ion{H}{ii}~regions, and we obtain the
total population by multiplying the result from Eq.~(\ref{eq:n_cl})
with the number of masers. In both cases the estimated number of
embedded stars is multiplied with
$\langle{}m_{\ast}\rangle{}$ to find the mass of the
embedded stellar population.

This total mass of the stellar population must be divided by some time
scale in order to obtain a star formation rate. In
Appendix~\ref{sec-app:sf-rate} we discuss that the association of
\ion{H}{ii}~regions and clouds must last for
$\tau_{\ion{H}{ii}}>1.1~{\rm{}Myr}$ to explain the number of
\ion{H}{ii} regions associated with clouds. This allows us to
calculate the star formation rate as
$\dot{M}_{\rm{}SF}=\langle{}m_{\ast}\rangle{}\cdot{}N_{\rm{}cl}/(1.1~{\rm{}Myr})$. This
is probably an upper limit to the true rate, given that we only have a
lower limit for $\tau_{\ion{H}{ii}}$.

We stress that the factors entering the calculation of the star
formation rate depend on the IMF. Specifically, the product
$\langle{}m_{\ast}\rangle{}\cdot{}N_{\rm{}cl}$ will be lower by a
factor 3.2 for given $M_{\rm{}max}$ if if we chose $\alpha_3=2.3$, a
value that resembles the \citet{salpeter1955:imf} result. The
resulting star formation rate would be lower by the same factor.

We essentially derive the star formation rate via multiplication of
some reference value with the number of \ion{H}{ii}~regions or
masers. The uncertainty in these sample sizes is governed by Poisson
statistics, as detailed in Appendix~\ref{sec-app:sf-rate}. Given the
small sample sizes, this typically yields uncertainties by a factor
two from the inferred SF rates. This representative uncertainty is
indicated in Fig.~\ref{fig:sf-mass}.

An independent estimate of the star formation rate can be obtained for
the $20~\rm{}km\,s^{-1}$ cloud. This region contains between 13 and 18
water masers \citep{lu2015:20kms}. In Appendix~\ref{sec-app:sf-rate}
we argue that these masers have lifetimes of order 0.16~Myr and are
powered by stars with mass $\gtrsim{}2.75\,M_{\sun}$. This yields
$\dot{M}_{\rm{}SF}=3\times{}10^{-3}\,M_{\sun}\,\rm{}yr^{-1}$.

Note that upper limits on the star formation rate are derived for
clouds that do not contain any evidence of embedded high--mass
stars. This follows from the fact that the non--detection of
high--mass stars sets an upper limit on $M_{\rm{}max}$ and thereby
$N_{\rm{}cl}$.

Also note that some very young high--mass stars might not manifest as
\ion{H}{ii}~regions since they are ``bloated'' and have low surface
temperatures \citep{hosokawa2009:bloated-stars}. This effect will not
significantly affect our estimates of star formation rates, however,
since we estimate the key parameter $\tau_{\ion{H}{ii}}$ directly from
the data.

\subsection{SF Suppression in CMZ
  Clouds\label{sec:sf-relation-comparison}}
Figure~\ref{fig:sf-mass} collects the star formation rates calculated
above. These rates are combined with Herschel--based cloud masses
above a column density threshold of
$9\times{}10^{22}~\rm{}cm^{-2}$. It is instructive to compare these
rates to values determined elsewhere in the Milky Way. Reference data
for this comparison come from \citet{lada2010:sf-efficiency}. They
show that molecular clouds within $\sim{}500~\rm{}pc$ from sun obey a
reference relation for star formation rate per unit dense gas of
\begin{equation}
\dot{M}_{\rm{}SF,ref} = (4.6\pm{}2.6)\times{}10^{-8}\,M_{\sun}\,{\rm{}yr}^{-1}
  \cdot{}(M_{\rm{}dense}/M_{\odot}) \, .
\label{eq:sf-reference}
\end{equation}
The mass of dense gas in this analysis, $M_{\rm{}dense}$, is taken to
be the cloud mass enclosed by an $\rm{}H_2$ column density threshold
of $6.7\times{}10^{21}~\rm{}cm^{-2}$ that corresponds to a visual
extinction $A_V=7~\rm{}mag$. The \citet{lada2010:sf-efficiency} data
and the relation describing these observations are shown in
Fig.~\ref{fig:sf-mass}. \citet{longmore2012:sfr-cmz} use WMAP radio
continuum maps obtain average star formation rates for several regions
in the CMZ, and they use observations of dust emission to
obtain measures of dense gas. These averages for three large regions
in the CMZ are also shown in Fig.~\ref{fig:sf-mass}.

Note that our Herschel--based mass estimates adopt a threshold of
$9\times{}10^{22}~\rm{}cm^{-2}$ to trace outer cloud boundaries. This
is much higher than the threshold of $6.7\times{}10^{21}~\rm{}cm^{-2}$
which \citet{lada2010:sf-efficiency} use to deliminate their
reservoirs of dense gas. The CMZ mass reservoirs of dense gas obtained
here should therefore be considered to be lower limits to the true
reservoir of dense gas. That said, we believe that very little gas
below the threshold of $9\times{}10^{22}~\rm{}cm^{-2}$ is directly
associated with the clouds considered here. It is plausible that the
estimates of the dense gas mass reservoirs are correct within a factor
$\ll{}10$.

Figure~\ref{fig:sf-mass} shows that a number of individual CMZ clouds
have star formation rates that are significantly below the reference
value set by $\dot{M}_{\rm{}SF,ref}$. As explained above, the dense
gas masses for CMZ clouds are lower limits. The lower mass limits
suggest that star formation is in some CMZ clouds suppressed by
factors $>10$, compared to the solar neighborhood. Other regions might
form stars at a rate consistent with $\dot{M}_{\rm{}SF,ref}$, such as
Sgr~C, Sgr~D, and Dust Ridge C. But this depends on how much the true
mass of dense gas in these regions deviates from the lower limit
measured here.

\section{Discussion\label{sec:discussion}}
\subsection{Linewidth--Size Relation,
  Shocks, and CMZ Cloud Orbits\label{sec:dissipation}}
Section~\ref{sec:linewidth-size} suggests that CMZ molecular clouds
appear to have an unusually steep linewidth--size relation, compared
to clouds elsewhere in the Milky Way. This raises the question why
such a trend might prevail in CMZ clouds.

It is plausible to speculate that this unusual linewidth--size trend
might be related to other peculiar features of CMZ cloud
kinematics. The latter includes unusually large line widths on spatial
scales $\gtrsim{}1~\rm{}pc$ and the prevalence of prominent shocks
that are for example evident in SiO emission. The shocks in
G0.253+0.016 are, for example, discussed by
\citet{kauffmann2013:g0.253}. \citet{jones2012:cmz-mopra} show that
most or all major CMZ molecular clouds are permeated by SiO
emission. Astrochemical models suggest that such SiO emission requires
shocks with relative velocities $\gtrsim{}20~\rm{}km\,s^{-1}$
\citep{guillet2009:shocks-sio}. The high line widths and shocks could
be the consequence of cloud--cloud collisions (e.g.,
\citealt{lis2001:ir-spectra} and \citealt{menten2009:g1.6}). The high
velocities that are apparently observed in these collisions could
reasonably result from the fast orbital motions of up to
$\sim{}200~\rm{}km\,s^{-1}$ in the CMZ \citep{molinari2011:cmz-ring,
  kruijssen2014:orbit} that can potentially lead to significant
velocity differences between clouds on slightly different orbits.

Other mechanisms that could raise the velocity dispersion include:
star formation feedback in the form of supernovae, stellar winds, and
radiation fields; occasional activity of $\rm{}Sgr~A^{\ast}$ in the
form of radiation and outflows; and the influence of $\sim{}10^5$
random main sequence stars that drift through a CMZ cloud at any given
moment\footnote{The stellar number density in the CMZ at given
  galactocentric radius $r$ is about
  $n_{\ast}=650~{\rm{}cm^{-3}}\cdot{}(r/{50~\rm{}pc})^{-0.8}$. This
  number can be estimated as
  follows. \citet{launhardt2002:cmz-potential} characterize the
  stellar mass distribution in the CMZ. \citet{kruijssen2014:orbit}
  show that this distribution can be approximated as
  $M=2\times{}10^8~M_{\sun}\cdot{}(r/{60~\rm{}pc})^{2.2}$. Differentiation
  of this relation provides the mass density as a function of $r$,
  while further division by the mean stellar mass
  $\langle{}m_{\ast}\rangle{}$ yields the stellar number density. A
  number of $\sim{}10^5$ in a cloud results if we then adopt a typical
  cloud radius of 3~pc at $r=50~\rm{}pc$, which is appropriate for our
  sample.}. These mechanisms must combine with the aforementioned
cloud--cloud collisions in a way that elevates the velocity
dispersions in clouds to the high observed levels.

The relatively narrow lines observed on spatial scales
$\lesssim{}0.1~\rm{}pc$ are, by contrast, consistent with expectations
for any molecular cloud. Recent work by
\citet{hacar2013:taurus-filament, hacar2015:musca} and
\citet{henshaw2014:g35} indicates that the dense star--forming
portions of clouds are composed of ``fibers'' with at most thermal
internal motions on spatial scales $\lesssim{}0.1~\rm{}pc$, i.e.,
$\sigma(v)\lesssim{}\sigma_{{\rm{}th},\langle{}m\rangle}(v)$ so that
$\mathcal{M}\lesssim{}1$. We thus would expect velocity dispersions
$\lesssim{}0.4~{\rm{}to}~0.6~\rm{}km\,s^{-1}$ on small spatial scales
in our target clouds, if we evaluate
$\sigma_{{\rm{}th},\langle{}m\rangle}(v)$ for gas temperatures between
50 and 100~K. This is roughly in line with our observations
(Sec.~\ref{sec:narrow-components}).

These plausible considerations give us an idea why the velocity
dispersion must decrease massively between large and small spatial
scales. The arguments do not tell us, though, why the velocity
dispersion must decrease \emph{steeply}. One can speculate that this
is a consequence of the unusually high velocity dispersions affecting
part of the clouds, in the sense that the dissipation of energy might
be particularly efficient at $\mathcal{M}\gg{}1$. We are not aware of
any theoretical argument supporting this case, though.

\subsection{SF is suppressed in Individual
  Clouds\label{sec:suppression-individual}}
The data summarized in Fig.~\ref{fig:sf-mass} have an obvious but
important consequence: CMZ star formation is --- at least in part ---
suppressed because \emph{individual CMZ clouds have suppressed
  SF}. This challenges models in which the low SF activity of the CMZ
purely results from the absence of dense and massive clouds. For
example, \citet{kruijssen2013:sf-suppression-cmz} summarize --- but do
not endorse --- models of this kind.  Here we show that numerous
massive and dense clouds do exists, but that many of these clouds are
for some reason not able to produce stars at the rate expected on the
basis of the cloud mass and average density.

This observation suggests that mechanisms suppressing SF in given
clouds are of particular importance for the CMZ. In
\citet{kauffmann2013:g0.253} we demonstrated that at least one CMZ
cloud is relatively devoid of massive dense cores of
$\lesssim{}0.1~\rm{}pc$ size that could form stars in an efficient
way. This trend has since been confirmed for the same cloud by, e.g.,
\citet{rathborne2014:g0253-pdf, rathborne2015:g0253-alma} and
\citet{johnston2014:g0.253}. \citetalias{kauffmann2016:gcms_ii} is largely dedicated to
exploring this point for the first time in a large comprehensive cloud
sample (i.e., the GCMS).

\subsection{Young Cloud Ages unlikely to explain Suppressed
  SF\label{sec:sgrb2_vs_g0253}}
One might speculate that a given CMZ molecular cloud evolves from an
early evolutionary state with little star formation to a later one
with high star formation activity. In that case the suppression of CMZ
star formation in dense gas might simply result from the fact that all
CMZ clouds are for some reason in an early evolutionary state before
the onset of star formation. We cannot here refute this model based on
a predominance of unevolved clouds. However, our data do show that the
evolution of SF activity between the clouds is more gradual than one
might think.

To give a specific example, here we show that the ``starless'' cloud
G0.253+0.016 and the ``mini--starburst'' Sgr~B2 might form about the
same number of stars per unit dense gas. Sgr~B2 contains at least 49
high--mass stars exciting \ion{H}{ii} regions, while no such object
exists in G0.253+0.016. However, Sgr~B2 contains about
$1.4\times{}10^6\,M_{\sun}$ of dense gas, while only
$9.3\times{}10^4\,M_{\sun}$ reside in G0.253+0.016
(Table~\ref{tab:target-summary}). Therefore, while the star formation
rate indicated by the number of \ion{H}{ii} regions differs by a
factor $\gtrsim{}49$ between the clouds, the mass of dense gas in
these regions also differs by a factor $\sim{}15$. In other words,
compared to G0.253+0.016 the Sgr~B2 region does indeed produce more
stars per unit dense gas --- but the difference established by the
observations of \ion{H}{ii} regions is only a factor
$\gtrsim{}49/15\approx{}3$.

This comparison suggests that the evolution of CMZ clouds with time
alone is unlikely to explain the low star formation activity in the
dense gas of the CMZ. Specifically, adjusting the mix between clouds
resembling Sgr~B2 and those similar to G0.253+0.016 might alter the SF
activity per unit dense gas merely by a factor $\sim{}3$: this number
might be significantly larger, given that we only have an upper limit
to the SF rate of G0.253+0.016. However, much larger differences in
the dense gas SF rates of clouds than observed now would be required
to explain the suppression of CMZ star formation in dense gas by a
factor $\gtrsim{}10$ \citep{longmore2012:sfr-cmz} via cloud evolution
alone.

One could still argue that \emph{all} CMZ clouds are in an early
evolutionary stage. This requires, however, synchronization between
the clouds to an unlikely extent. Consider the evolution of individual
clouds. It is plausible to think that these evolve towards efficient
star formation on a time scale resembling few multiples of the
free--fall
($3\times{}10^5~{\rm{}yr}\cdot{}[n({\rm{}H_2})/10^4~{\rm{}cm^{-3}}]^{-1/2}$;
\citealt{kauffmann2008:mambo-spitzer}) and crossing time scales
($\sim{}3\times{}10^5~{\rm{}yr}$;
Sec.~\ref{sec:linewidth-size}). \emph{All} CMZ clouds must be younger
than this if they feature little SF due to a low age, and so their
evolution must be synchronized to on a time scale smaller than their
age, $\lesssim{}3\times{}10^5~\rm{}yr$. It is hard to see how this
should be achieved: disturbances probably travel along the CMZ cloud
orbit on a time scale resembling the orbital period
$\gtrsim{}4~\rm{}Myr$.  Such a synchronization appears unlikely,
unless it is induced by powerful processes affecting the entire
CMZ. Activity in $\rm{}Sgr~A^{\ast}$ would be a possible candidate for
such far--reaching influences.

\section{Summary\label{sec:summary}}
This paper introduces the Galactic Center Molecular Cloud Survey
(GCMS), the first systematic study resolving all major molecular
clouds in the Central Molecular Zone (CMZ) of the Milky Way at
interferometer angular resolution. Here we present images of the
$\rm{}N_2H^+$ (3--2) transition and dust continuum emission near
280~GHz frequency. These are produced by combining data from the
Submillimeter Array (SMA) and the Atacama Pathfinder Experiment
(APEX). We also include dust continuum images from the Herschel Space
Telescope at wavelengths between 160 and $500~\rm{}\mu{}m$, as well as
an extensive literature survey for star formation associated with CMZ
molecular clouds.

A companion paper (\citealt{kauffmann2016:gcms_ii}; hereafter
\citetalias{kauffmann2016:gcms_ii}) uses the dust continuum images to
study the density structure of CMZ clouds. Here we describe the sample
and the observations (Sec.~\ref{sec:observations}), study the cloud
kinematics (Sec.~\ref{sec:kinematics}), and research the star
formation (SF) efficiency in the dense gas of individual CMZ molecular
clouds (Sec.~\ref{sec:suppression-sf}). This leads to the following
conclusions.
\begin{itemize}
\item We present the first analysis of the linewidth--size relation of
  CMZ clouds down to spatial scales $\sim{}0.1~\rm{}pc$. High velocity
  dispersions $\ge{}6.5~\rm{}km\,s^{-1}$ observed for cloud radii
  $\ge{}1~\rm{}pc$ reduce to velocity dispersions of 0.6 to
  $2.2~\rm{}km\,s^{-1}$ on spatial scales $\sim{}0.1~\rm{}pc$
  (Sec.~\ref{sec:linewidth-size}, Fig.~\ref{fig:linewidth-size}). This
  implies an unusually steep linewidth--size relation
  $\sigma(v)\propto{}r_{\rm{}eff}^{0.66\pm{}0.18}$, i.e., with a power
  law index that appears to be significantly larger than elsewhere in
  the Milky Way. This steep slope might be the result of the decay of
  fast orbit--induced large--scale motions to transonic conditions on
  small spatial scales (Sec.~\ref{sec:dissipation}).
\item Studies of the density structure and star formation activity of
  CMZ clouds (Secs.~\ref{sec:suppression-sf},
  Appendices~\ref{sec-app:sf-observations} and \ref{sec-app:sf-rate})
  show that the star formation activity in \emph{individual} CMZ
  clouds is suppressed by an order of magnitude
  (Sec.~\ref{sec:suppression-individual}) relative to star formation
  relations that apply to the solar neighborhood (e.g.,
  \citealt{heiderman2010:sf-law, lada2010:sf-efficiency}). This
  eliminates models that try to explain the suppression of CMZ star
  formation purely via a suppression of the formation of dense
  molecular clouds.
\item There is no evidence to think that CMZ star formation is
  suppressed because a large number of CMZ clouds reside in an early
  evolutionary phase before the onset of significant SF
  (Sec.~\ref{sec:sgrb2_vs_g0253}). For example, the star formation
  rate per unit dense gas might differ by only a
  factor $\sim{}3$ between the prototypical ``starless'' cloud
  G0.253+0.016 and the ``mini--starburst'' in Sgr~B2. Such variations
  between clouds, that could plausibly result from cloud evolution,
  are not sufficient to explain why CMZ star formation in dense gas is
  suppressed by a factor $\gtrsim{}10$.
\end{itemize}

\begin{acknowledgements}
  We thank a thoughtful and helpful referee who provided a thorough
  review full of insights. We also greatly appreciate help from
  Diederik Kruijssen and Steve Longmore who provided detailed feedback
  on draft versions of the paper. These combined sets of comments
  helped to significantly improve the quality and readability of the
  paper before and during the refereeing process. This research made
  use of astrodendro, a Python package to compute dendrograms of
  astronomical data (\path{http://www.dendrograms.org/}). JK and TP
  thank Nissim Kanekar for initiating a generous invitation to the
  National Centre for Radio Astrophysics (NCRA) in Pune, India, where
  much of this paper was written. QZ acknowledges support of the
  SI~CGPS grant on Star Formation in the Central Molecular Zone of the
  Milky Way.This research was carried out in part at the Jet
  Propulsion Laboratory, which is operated for the National
  Aeronautics and Space Administration (NASA) by the California
  Institute of Technology. AEG acknowledges support from NASA grants
  NNX12AI55G, NNX10AD68G and FONDECYT grant 3150570.
\end{acknowledgements}

\bibliographystyle{aa}
\bibliography{/Users/jens/texinputs/mendeley/library}

\Online

\begin{appendix}

\section{Imaging of Interferometer Data\label{sec-app:imaging}}
Our project is made complex by the need to image extended emission
with the interferometer. The mathematics of so--called zero--spacing
corrections are well understood. For example, \citet{koda2011:m51}
provide a complete imaging pipeline for this problem. This pipeline
has inspired much of the processing done below, but it requires a
customized version of MIRIAD. We are not aware of a well--documented
and easily useable implementation of this procedure. We therefore
continue to collect and document procedures at
\begin{equation*}
\text{\url{http://tinyurl.com/zero-spacing}}\quad{}.
\end{equation*}
Other relevant discussions of this problem are provided by
\citet{kurono2009:zero-spacing}, and the IRAM technical
report\footnote{accessible at \url{http://www.iram-institute.org}}
2008--2 by Rodriguez-Fernandez, Pety \& Gueth.

\subsection{Imaging of Dust Emission\label{sec-app:imaging-dust}}
We use CASA for the imaging of dust continuum emission. We start with
the spectrally averaged data produced in the MIRIAD processing laid
out in Sec~\ref{sec:obs-sma-apex}. The data are then imaged and
cleaned. We use a pixel size of $0\farcs{4}$ for all sources, which
resolves the synthesized beams well. Multi--scale clean is employed:
we use scales of $(0,5,10,15,20)$ pixels. Briggs weights of 0.5 are
chosen. The cleaning threshold is set to $6~\rm{}mJy\,beam^{-1}$,
which is at the upper end of the image noise levels we derive
below. Up to 5,000 cleaning iterations are used. The data are jointly
imaged and cleaned as a mosaic.

We use an iterative scheme to image and clean the data. The main goal
is to reduce the presence of obvious artifacts. The presence of such
structures, for example in the form of periodic ripples, is hard to
quantify. We therefore optimize the scheme via visual inspection of
the resulting maps. This in particular includes the number of
iterations after which no significant change is obvious, and the range
of angular scales for multi--scale cleaning.

In the first step of our iteration scheme we clean the entire image
down to the aforementioned threshold. This image is used to obtain a
model for further iterations. To do this we highlight obvious emission
features in the map using a generously drawn polygon. A model is then
created from the resulting image by setting the emission outside of
this boundary, as well as all negative features, to zero. Four more
iteration steps follow the same procedure, with the exception that
cleaning is constrained to the polygon drawn around significant
emission and that CASA's CLEAN uses an emission model provided via the
MODELIMAGE parameter. The procedure also creates a gain image (suffix
``flux'' in CASA terminology) that encodes how sensitive the final
image is to sources in the outer parts of the map.

In a final step we combine the image generated from the SMA data with
an estimate of the extended dust emission filtered out by the
interferometer data reduction. This correction uses CASA's FEATHER
tool. It is based on observations from the APEX Telescope Large Area
Survey of the Galaxy (ATLASGAL, \citealt{schuller2009:atlasgal}; see
\citealt{contreras2013:atlasgal-catalogue} for data reduction details)
that were done using the Large APEX Bolometer Camera (LABOCA;
\citealt{siringo2007:laboca,siringo2009:laboca}). The LABOCA
observations were done at a wavelength of $870~\rm{}\mu{}m$ with a
beam size of $19\farcs{}2$. We follow the formalism of
\citet{kauffmann2008:mambo-spitzer} to scale the dust emission to
280~GHz, the frequency of the SMA observations, by assuming a dust
temperature of 20~K and a dust opacity varying with frequency as
$\nu^{\beta}$ with $\beta=1.75$ (taken from
\citealt{battersby2011:cluster-precursors} based on
\citealt{ossenkopf1994:opacities}). In practice we therefore multiply
the LABOCA maps by a factor 0.50 to do this conversion. The predicted
dust emission map is then multiplied by the aforementioned gain
image. This makes sure that both the processed single--dish file and
the interferometer--only data only contain data on emission from the
same area of the sky\footnote{We thank Thomas Stanke from the European
  Southern Observatory (ESO) for making this suggestion.}. In a final
preparatory step we make sure that the header of the processed LABOCA
map states that the beam has a size equal to the original value, i.e.\
$19\farcs{}2$. FEATHER is then run on the interferometer--only map and
the processed LABOCA image. We enable the ``lowpassfiltersd'' option
in this step.

Note that the scaling of the LABOCA map to the observing frequency of
the SMA introduces negligible uncertainties $\lesssim{}5\%$ in this
procedure. This dust opacity law from
\citet{battersby2011:cluster-precursors} approximates tabulated values
from \citet{ossenkopf1994:opacities}. We note that
\citet{rathborne2014:g0253-pdf} recently commented that
$\beta\sim{}1.2~{\rm{}to}~1.5$, based on dust emission spectral energy
distributions from Herschel, but they also suggest that
$1\le{}\beta{}\le{}2$ is broadly consistent with the data. A detailed
study is needed to bring clarity. However, for a dust temperature of
20~K, experimentation shows that the scaling factor between dust
intensities at $870~\rm{}\mu{}m$ and 280~GHz deviates by $\le{}5\%$
relative to our choice of $\beta=1.75$ in the range
$1.5\le{}\beta{}\le{}2.0$.  Our processing of Herschel data suggests
dust temperatures in the range 18 to 25~K for our targets (Sec.~\ref{sec:obs-herschel}),
which is similar to the 20~K \citet{longmore2011:m025} find for
G0.253+0.016. Provided $\beta=1.75$, the scaling factor deviates by
$\le{}3\%$ relative to our reference choice of 20~K for dust
temperatures in the range of 15 to 25~K. More significant
uncertainties in imaging can arise due to calibration uncertainties of
SMA and APEX data, as well as the fundamental assumptions made within
FEATHER.

\subsection{Imaging of Line Emission\label{sec-app:imaging-lines}}
We combine actual interferometer observations with synthetic data
derived from single--dish observations to populate the entire
$uv$--plane from its center out to the highest spatial frequencies
probed by the interferometer. This requires a number of processing
steps which we implement in MIRIAD; all of the tasks mentioned below
are part of this package. Below we concentrate on problems with
imaging and cleaning. However, we warn that much of the development
work actually concerns the accurate transfer of information between
MIRIAD tasks. Careful testing is crucial to guarantee that assumptions
made within MIRIAD are properly taken into account. A pixel size of
$1\arcsec$ is used for all maps, which is sufficient to resolve the
synthesized beam size.

We first use UVAVER to resample the interferometer into specific
velocity channels. A dirty map of the entire interferometer data set
is produced. The single--dish data are first smoothed to the spectral
resolution of the interferometer data and then resampled into the
frequency channels of the interferometer data. The
resulting single--dish data are scaled to the $T_{\rm{}mb}$--scale
using the efficiencies listed on the APEX
website\footnote{\url{http://www.apex-telescope.org/telescope/efficiency/index.php}}.

The processed single--dish data are then cleaned using a Gaussian
model approximating the beam of the single--dish telescope. The idea
behind this step is as follows. Many structures in single--dish maps
are actually rather compact features that get blurred by the beam of
the single--dish telescope. We need to identify these features since
the interferometer will also detect them. We also must make sure that,
as far as possible, the single--dish data does not contain corrupted
--- i.e., blurred --- information on these compact
structures. Careless processing can otherwise have the result that
information on the same compact source is entering into the data
reduction twice --- i.e., once uncorrupted via the interferometer
observations, and once corrupted via the blurred single--dish data. We
use the \citet{steer1984:clean} cleaning algorithm with a gain of 0.01 and
a clipping level of 0.85.

To calculate $uv$--data from the single--dish observations we first
pick $10^4$ points in the $uv$--plane via randomized drawing. The
density distribution of these points follows the Gaussian distribution
resulting from a Fourier transform of the Gaussian pattern with
$23\farcs{}7$ width at half power that is assumed to describe the
resolution of the single--dish map. Data points in the $uv$--plane
representing baselines beyond 80\% of the telescope diameter are
further explicitly removed. We then loop through the pointing centers
of the interferometer mosaic. For every position we generate a map
representing the primary beam sensitivity pattern of the
interferometer observations. The single--dish map is then multiplied
with this sensitivity distribution. This processed single--dish image
is then used to generate synthetic visibilities at the randomized
$uv$--plane sample points generated above. Experiments show that
imaging of these synthetic $uv$--data recovers the original
single--dish map.

We then produce a combined set of $uv$--data that contains information
from both the interferometer and single--dish observations. One critical
aspect is that we must down--weight the $\sim{}10^4$ single--dish data
points in order not to suppress the interferometer data. We implement
this by artificially increasing the system temperature of the
single--dish data. The scaling factor is obtained as described further
below. The the actual data combination and weighting is performed in
MIRIAD's INVERT task with the ``systemp'' weighting option enabled. We
use up to $10^5$ iterations of \citet{steer1984:clean} cleaning with a
gain of 0.01 and a clipping level of 0.85 to deconvolve the images.

We vary the weighting factor between the single--dish and
interferometer data to optimize the size and shape for the dirty
beam. It is hard to devise quantitative criteria to do this
adjustment. We manually minimize sidelobes and select small beams via
manual inspection of the dirty beam.

At this point it is critical to realize that (\textit{i})~cleaning of
extended emission leaves large residuals not described by clean
components, and (\textit{ii})~the signal in these residuals, expressed
in the $\rm{}Jy\,beam^{-1}$--scale, is calibrated to the area of the
dirty beam \emph{which is generally different from the area of the
  restoring beam}. Item~(\textit{i}) means that we must combine the
residual with the clean components to produce a map that captures all
spatial scales contained in the data. Item~(\textit{ii}) means that we
must do this carefully, i.e., by scaling the data to a common
calibration of the $\rm{}Jy\,beam^{-1}$--scale. We do the latter by
calibrating the residual, obtained via MIRIAD's RESTOR task run in
``residual'' mode, to the area of the restoring beam. First, a
representative dirty beam is produced via the MOSPSF task. Second, the
area contained in this beam is measured. Third, the residual is
multiplied by the ratio between the area of the restore beam to the
area of the dirty beam. Fourth, the clean components are convolved
with the restore beam. In a final step the convolved clean components
are added to the re--calibrated residual map to obtain the final data
set. We find that this procedure guarantees that the resulting map
recovers the original intensities to $<10\%$ of the values seen by the
single--dish telescope.

\section{Observations of Star Formation in CMZ Clouds
  \label{sec-app:sf-observations}}
\subsection{Sgr~C}
\citet{liszt1995:sgr_c} use a 1.6~GHz survey of the CMZ by
\citet{liszt1992:sgr_survey} to characterize the emission of
Sgr~C. The emission from this region is found to be
thermal. H70$\alpha$ recombination lines yield a velocity of
$-65~\rm{}km\,s^{-1}$. The continuum emission is rather extended, with
a radius of $1\arcmin$ corresponding to 2.3~pc at the distance of the
CMZ; this is not a compact \ion{H}{ii} region. A single star of
spectral type O4 or O6 could provide the required ionization.

The \ion{H}{ii} region is not directly embedded in the dust cloud as
here extracted from Herschel maps. This region of $1\arcmin$ (i.e.,
$\sim{}2~\rm{}pc$) radius is outside of the Herschel--derived cloud
boundaries by at least $1\arcmin$. The Sgr~C cloud appears to reside
exactly at the boundary of the \ion{H}{ii} region and might have been
sculpted by it. This would indicate that the molecular cloud and the
\ion{H}{ii} region are indeed connected. For example, the cloud as
seen today might be a leftover of the star formation process.
However, there is no clear indication of a direct link, such as
coincidence within the Herschel--derived boundaries. We abstain from
claiming a speculative association between this large \ion{H}{ii}
region and the Sgr~C cloud.

Instead we build on a detailed study of star formation in
Sgr~C. \citet{kendrew2013:sgr-c} demonstrate that a compact and faint
\ion{H}{ii} region initially detected by
\citet{forster2000:radio-continuum} is embedded in this cloud. The
radio data have not been used to determine a spectral type: here we
adopt a mass of $18\,M_{\sun}$ to be conservative.

\subsection{\sansmath{}$20~km\,s^{-1}$ Cloud}
The thermal radio source ``G'' of \citet{ho1985:cmz_radio} resides
well inside the boundaries of the $20~\rm{}km\,s^{-1}$ cloud. It is
interpreted as a single \ion{H}{ii} region or a cluster of such
sources. A single star of spectral type O9
could excite this region. Further, \citet{lu2015:20kms}
report the discovery of up to 18 water masers in this cloud.

\subsection{\sansmath{}$50~km\,s^{-1}$ Cloud}
\citet{mills2011:continuum-50kms} provide a recent summary of this
region. A chain of four previously known \ion{H}{ii} regions, labeled
``A'' through ``D'' from north to south, is detected. The most
southern region ``D'' is the most compact and presumably youngest
object. The spectral types of the exciting stars listed in
Table~\ref{tab:sf-tracers} are derived assuming a single ionizing
source. \citet{yusef-zadeh2010:50kms} use observations of \ion{Ne}{ii}
to derive systemic velocities $\sim{}40~\rm{}km\,s^{-1}$ for the
targets.

\subsection{G0.253+0.016}
The star formation status of this region is
controversial. \citet{immer2012:multi-wavelength-cmz} find no radio
continuum source inside this cloud. Their source ``A'' lies
$1\farcm{}3$ outside of the Herschel--derived cloud boundary. The
exciting source could be a single O9 star. There is no obvious reason
to think that source ``A'' from
\citet{immer2012:multi-wavelength-cmz} was recently born in
G0.253+0.016.

However, using the same data, \citet{rodriguez2013:g0.253} find
several additional sources in the vicinity of G0.253+0.016. Their
source ``3'' is the object ``A'' discussed by
\citet{immer2012:multi-wavelength-cmz}. A radial velocity of
$45~\rm{}km\,s^{-1}$ from previous recombination line observations is
quoted. Sources ``1'', ``2'', and ``7'' have non--thermal spectral
indexes indicating that these are not \ion{H}{ii} regions. Sources
``4'', ``5'', and ``6'' have a thermal nature. Their flux densities
are each consistent with excitation by a single B0.5 star. These
objects are inside the cloud boundary defined via the Herschel data,
except for source ``6'' which is offset by $0\farcm{}2$.  Given the
sensitivities achieved by the observations, all of the targets seen by
\citet{rodriguez2013:g0.253} should also have been detected by
\citet{immer2012:multi-wavelength-cmz}. The most plausible explanation
for this discrepancy is that \citeauthor{rodriguez2013:g0.253} detect
an extended component of radio continuum emission which is not picked
up in the data reduction of
\citeauthor{immer2012:multi-wavelength-cmz} This is in particular
suggested by such emission being detected in new VLA maps by
\citet{mills2015:cmz-masers}. Extended radio emission would not hint
at star formation in G0.253+0.016 but rather would reflect the general
excitation conditions in the CMZ. We therefore ignore the three
sources found by \citet{rodriguez2013:g0.253}.

\subsection{Dust Ridge ``C'' \& ``D''}
These regions are surveyed by
\citet{immer2012:multi-wavelength-cmz}. No radio sources associated
with these clouds are detected.

\subsection{Sgr~B1--off}
\citet{mehringer1993:sgr_b1} find a number of compact sources in the
vicinity of Sgr~B1--off. Their sources ``A'' ($0\farcm{}5$ offset, i.e.\
1.2~pc) and ``C'' ($0\farcm{}9$ offset, i.e.\ 2.1~pc) are those
closest to the dust emission feature we extract from Herschel
maps. Following \citet{mehringer1993:sgr_b1}, single stars with
spectral types B0 would be sufficient to excite these
regions. However, there is no obvious reason to think that these
\ion{H}{ii} regions are physically associated with Sgr~B1--off.

\subsection{Sgr~B2}
\citet{gaume1995:sgr_b2-i} detect a total of 49 compact radio
continuum sources in this region. The numbers given in
Table~\ref{tab:sf-tracers} assume that every source is excited by a
single star.

\subsection{Sgr~D}
\citet{liszt1992:sgr_survey} finds two structures in Sgr~D. One is a
supernova remnant (SNR). $\rm{}H_2CO$ absorption lines at up to
$+125~\rm{}km\,s^{-1}$ are seen against this source. The other one is
an \ion{H}{ii} region for which H70$\alpha$ recombination lines yield
a velocity of $-19~\rm{}km\,s^{-1}$. However, the aforementioned
$\rm{}H_2CO$ lines are also seen against this source. The \ion{H}{ii}
region contains a compact core to which \citet{liszt1992:sgr_survey}
refers as source ``3''. He derives an ionization rate consistent with
a single O7 star for this core, if the source is located in the
CMZ. Similar results were derived by \citet{mehringer1998:sgr_d}. This
source falls directly on top of the features in our SMA maps. It is
clearly associated with the Sgr~D molecular cloud.

\section{Estimates of the Star Formation Rate\label{sec-app:sf-rate}}
The observations of star formation reviewed above only constrain the
presence of high--mass stars in the target clouds. But the presence
--- or absence --- of high--mass stars hints at the total stellar
population embedded in the clouds, since the relative number of stars
of low and high mass are related via the initial mass function
(IMF). In this study we adopt the \citet{kroupa2002:imf} IMF with
$\alpha_3=2.7$. Consider a stellar cluster that is described by this
IMF. Fitting his results with a power law shows that the number of
cluster members down to a mass of $0.01\,M_{\sun}$, $N_{\rm{}cl}$,
depends on the mass of the most massive star in the cluster,
$M_{\rm{}max}$, as
\begin{equation}
N_{\rm{}cl} \approx 20.5 \cdot (M_{\rm{}max} / M_{\sun})^{1.7} \, .
\label{app-eq:n_cl}
\end{equation}
We adopt the mean stellar mass of $0.29\,M_{\sun}$ that holds for the
\citet{kroupa2002:imf} IMF with
$\alpha_3=2.7$. Equation~(\ref{app-eq:n_cl}) permits to estimate
masses for the stellar populations associated with the CMZ clouds. We
now need to divide these masses by a timescale in order to obtain a
star formation rate.

\subsection{Lifetimes of H\,\textsc{ii} Regions in the CMZ}
Following \citet{weidner2010:o-star-masses}, coeval main
sequence stars can populate the spectral types of relevance here
(i.e., up to O4) for $2~{\rm{}to}~7~\rm{}Myr$. One could use these
numbers if it were certain that the stars power \ion{H}{ii} regions
throughout their lifetime. But this is not clear, in particular not in
a region like the CMZ.

For example, the high velocity dispersions inside the clouds might
mean that the stars produced in these regions are born with high
relative velocities with respect to their natal clouds. In that case
the stars powering the \ion{H}{ii} regions might quickly detach from
their birth sites. If present, this process might occur on time scales
$\gtrsim{}t_{\rm{}cross}$, provided that velocity also controls the
relative motion of stars and gas inside a cloud. The discussion in
Sec.~\ref{sec:linewidth-size} gives crossing times $>0.3~\rm{}Myr$ on
intermediate to large spatial scales.

Alternatively we can combine the global star formation record of the
CMZ with the observed number of \ion{H}{ii} regions to constrain the
timescale over which \ion{H}{ii} regions are associated with CMZ
molecular clouds. \citet{longmore2012:sfr-cmz} obtain a star formation
rate of $<0.06\,M_{\sun}\,\rm{}yr^{-1}$ for the region
$|\ell{}|\le{}1\degr$ using free--free foregrounds extracted from the
WMAP data, when considering all radiation at $|b|\le{}1\degr$. Here we
consider this to be an upper limit since it is likely that radiation
from $\rm{}Sgr~A^{\ast}$ unrelated to the star formation process
contributes to the ionizing radiation characterized in this
experiment. \citet{longmore2012:sfr-cmz} use results from
\citet{murray2010:wmap-sfr} to estimate that the high--mass stars
traced by the free--free maps have a mean age
$\tau_{\rm{}HM}=4~\rm{}Myr$. Now we use Eq.~(\ref{app-eq:n_cl}) to
find that 2,791 stars are needed to produce one high--mass star of
$18\,M_{\sun}$ that can power an \ion{H}{ii} region. This number can
be multiplied with the mean stellar mass of $0.29\,M_{\sun}$ and
divided by the timescale $4~\rm{}Myr$ to find that a star formation
rate of $2\times{}10^{-4}\,M_{\sun}\,{\rm{}yr^{-1}}$ is required to
produce one star of a mass high enough to power an \ion{H}{ii}
region. This suggests that
$N_{\rm{}HM}<0.06\,M_{\sun}\,{\rm{}yr^{-1}}/(2\times{}10^{-4}\,M_{\sun}\,{\rm{}yr^{-1}})\approx{}300$
such stars exist in the inner CMZ where $|\ell{}|\le{}1\degr$. This
count can be compared to the number of known \ion{H}{ii} regions in
the same volume: combining data from \citet{ho1985:cmz_radio},
\citet{mehringer1993:sgr_b1}, \citet{liszt1995:sgr_c},
\citet{gaume1995:sgr_b2-i}, \citet{mills2011:continuum-50kms}, and
\citet{immer2012:multi-wavelength-cmz} we find
$N_{\ion{H}{ii}}\ge{}80$ such objects at $|\ell|\le{}1\degr$. We can
now make the assumption that the relative number ratio between the
\ion{H}{ii} regions and the total count of high--mass stars is equal
to the ratio between the duration of the phase where \ion{H}{ii}
regions are embedded in clouds and the mean lifetime of high--mass
stars,
$N_{\ion{H}{ii}}/N_{\rm{}HM}\sim{}\tau_{\ion{H}{ii}}/\tau_{\rm{}HM}$. Rearrangement
and substitution gives $\tau_{\ion{H}{ii}}>1.1~{\rm{}Myr}$. We
consider this a lower limit because of the aforementioned contribution
of $\rm{}Sgr~A^{\ast}$ to the ionizing radiation and the lower limit
on the number of \ion{H}{ii} regions (from our non--systematic
literature survey).

We adopt this number of $1.1~{\rm{}Myr}$, which is also consistent
with the qualitative estimate
$\tau_{\ion{H}{ii}}\gtrsim{}t_{\rm{}cross}>0.3~\rm{}Myr$ obtained
above. This value is also broadly consistent with the age of the
stellar population that is immediately associated with the Orion
Nebula \citep{bally2008:orion}. Substitution yields
$\dot{M}_{\rm{}SF}=0.29\,M_{\sun}\cdot{}N_{\rm{}cl}/(1.1~{\rm{}Myr})$.

\subsection{Reference Relations for Star Formation}
A reference dense gas star formation rate that holds for
clouds near the Sun is provided by \citet{lada2010:sf-efficiency},
\begin{equation}
\dot{M}_{\rm{}SF,ref} = (4.6\pm{}2.6) \times 10^{-8} \, M_{\sun} \,{\rm{}yr}^{-1}
   \cdot (M_{\rm{}dense}/M_{\sun}) \, .
\label{eq-app:sf-ref}
\end{equation}
It is obtained by multiplication of the number of embedded young
stellar objects, $N_{\rm{}YSO}$, with a mean stellar mass of
$0.5\,M_{\sun}$ and division by a maximum age of young stellar objects
of 2~Myr, so that
$\dot{M}_{\rm{}SF}=0.5\,M_{\sun}\cdot{}N_{\rm{}YSO}/(2~{\rm{}Myr})$.  In
\citet{kauffmann2013:g0.253} we argued that this value should be
increased by some factor because infrared observations only reveal a
fraction of the true population of embedded young stars. Here we make the more
conservative assumption that all embedded stars are counted, and we
follow \citet{evans2009:c2d-summary, evans2014:sfr-nearby-clouds} in
adopting a mean stellar mass of $0.5\,M_{\sun}$, equal to that of
well--studied populations of embedded stars. Increases in
$\dot{M}_{\rm{}SF,ref}$, as appropriate if a significant number of
stars are missed, would imply that CMZ star formation is suppressed by
factors larger than those obtained in the main text.

We note that some of the assumptions made in the analysis of
\ion{H}{ii} regions are not entirely consistent with those made in the
analysis of infrared samples of embedded young stars. For example,
both \citet{evans2009:c2d-summary, evans2014:sfr-nearby-clouds} and
\citet{lada2010:sf-efficiency} assume a mean stellar mass of
$0.5\,M_{\sun}$ where we adopt a value of $0.29\,M_{\sun}$. This
difference comes about because infrared surveys are not complete down
to the substellar regime, while the \citet{kroupa2002:imf} IMF
analysis conducted above considers masses down to
$0.01\,M_{\sun}$. For example, the mean stellar mass of the
\citet{kroupa2002:imf} IMF for stellar masses $>0.1\,M_{\sun}$ is
$0.5\,M_{\sun}$, consistent with what is adopted for the analysis of
infrared surveys. However, only 8\% of the stellar mass in the
\citet{kroupa2002:imf} IMF resides in the mass interval of
$0.01~{\rm{}to}~0.1\,M_{\sun}$. One would need to lower the CMZ star
formation rates by this percentage to achieve full conceptual
consistency. Here we omit this small correction due to the other
systematic uncertainties involved.

\subsection{Data on H\,\textsc{ii} Regions}
The spectral types of the sources exciting \ion{H}{ii}~regions
indicate the masses of the illuminating stars. We adopt the mean
initial mass as a function of the spectral type for dwarf stars with
realistic rotation from Table~4 of \citet{weidner2010:o-star-masses}
to do this calculation.

\ion{H}{ii}~regions are powered by stars of spectral type O9.5 or
earlier. These stars have masses $>18\,M_{\sun}$ following
\citeauthor{weidner2010:o-star-masses}. 
Equation~(\ref{app-eq:n_cl}) can be interpreted in the sense that on
average a group of 3,059 stars is needed to form one star of
$\ge{}19\,M_{\sun}$. Following
$\dot{M}_{\rm{}SF}=0.29\,M_{\sun}\cdot{}N_{\rm{}cl}/(1.1~{\rm{}Myr})$
such a population implies a star formation rate of
$8\times{}10^{-4}\,M_{\sun}\,{\rm{}yr}^{-1}$. The presence of
$N_{\ion{H}{ii}}$ \ion{H}{ii}~regions thus statistically implies a
star formation rate
\begin{equation}
\dot{M}_{\rm{}SF} = 8\times{}10^{-4}\,M_{\sun}\,{\rm{}yr}^{-1} \cdot
N_{\ion{H}{ii}} \, .
\end{equation}
We use this relation to derive star formation rates from the
observational data on \ion{H}{ii}~regions. This approach suffers from
statistical uncertainties due to how actual star formation in a region
samples the IMF. Provided $N_{\rm{}cl}$ is held constant, deviations
between the statistically expected number of \ion{H}{ii}~regions and
the actual one are described by Poisson statistics. For these the
standard deviation of a measurement with expectation value $x$ is
given by $\sigma(x)=x^{1/2}$. This suggests to use
$\sigma(N_{\ion{H}{ii}})=N_{\ion{H}{ii}}^{1/2}$ and
\begin{equation}
\sigma(\dot{M}_{\rm{}SF}) = 8\times{}10^{-4}\,M_{\sun}\,{\rm{}yr}^{-1} \cdot
N_{\ion{H}{ii}}^{1/2}
\end{equation}
to estimate the \emph{statistical} uncertainties on
$\dot{M}_{\rm{}SF}$. This implies uncertainties of order of a factor 2
for the small numbers of \ion{H}{ii}~regions detected in most of our
targets. This is the uncertainty indicated in
Fig.~\ref{fig:sf-mass}. A much lower relative factor applies in Sgr~B2
due to the rich sample of high--mass stars (the absolute value of the
uncertainty increases, though). On top of these there are further
\emph{systematic} uncertainties, e.g., due to uncertainties in the
shape of the IMF and the relationship between stellar masses and
spectral types. An assessment of these possible sources of error is
well beyond the scope of the current study.

Alternatively one could use the mass of the star with the earliest
spectral type to obtain $N_{\rm{}cl}$ from Eq.~(\ref{app-eq:n_cl}) and
use this number to derive the SF rate. However, the maximum mass
sampled from an IMF can deviate from the expectation value by factors
$\sim{}2$ (see black lines in Fig.~1 of
\citealt{kruijssen2012:globular-clusters} for sampling from a
power--law distribution function, based on a formalism by
\citealt{maschberger2008:cluster-membership}). Unfortunately,
$N_{\rm{}cl}$ strongly depends on $M_{\rm{}max}$ and the mass
uncertainty implies uncertainties by factors $\sim{}2^{1.7}=3.25$
relative to the expectation value\footnote{We thank D.~Kruijssen for
  kindly pointing out this problem.}. While potentially useful, here
we abstain from such estimates.\medskip

\noindent{}We use data on the Gould Belt molecular clouds in the solar
neighborhood (i.e., $d\lesssim{}500~\rm{}$) from
\citet{lada2010:sf-efficiency} to conduct a consistency check of our
method. Except for Orion, none of the Gould Belt clouds contain
\ion{H}{ii} regions. This implies stellar masses $\le{}18\,M_{\sun}$
and therefore
$\dot{M}_{\rm{}SF}<0.8\times{}10^{-3}\,M_{\sun}\,\rm{}yr^{-1}$. The
latter star formation rate is in line with the data compiled by
\citet{lada2010:sf-efficiency}.

In Orion we do find high--mass stars and a major \ion{H}{ii}
region. \citet{hillenbrand1997:orion} identify five stars of mass
$\ge{}20\,M_{\sun}$ in this region.  In this case we obtain
$N_{\rm{}cl}$ for $M_{\rm{}max}=20\,M_{\sun}$ and increase that value
by a factor 5 to derive the SF rate from
$\dot{M}_{\rm{}SF}=0.29\,M_{\sun}\cdot{}N_{\rm{}cl}/(1.1~{\rm{}Myr})$. We
find
$\dot{M}_{\rm{}SF}=4.4\times{}10^{-3}\,M_{\sun}\,\rm{}yr^{-1}$. This
is a bit high compared to the rate of
$7\times{}10^{-4}\,M_{\sun}\,\rm{}yr^{-1}$ adopted by
\citet{lada2010:sf-efficiency}.

\subsection{Data on Methanol Masers}
Here we assume that class~II methanol masers indicate the formation of
stars of $\sim{}20\,M_{\sun}$ mass. The 6.7~GHz masers used here are
almost exclusively associated with luminous cloud fragments that are
massive enough to form high--mass stars \citep{urquhart2014:masers}
--- but this finding gives no clear minimum stellar mass for their
excitation. Here we adopt a rather high mass to be conservative (i.e.,
obtain upper limits to the star formation rate).

Two of our target clouds host such masers but are devoid of
\ion{H}{ii} regions. This might be the case because these high--mass
stars are still deeply enshrouded in the cloud and are unable to
ionize a detectable volume of gas. The star formation rate listed in
Table~\ref{tab:sf-tracers} is calculated assuming the aforementioned
value as the maximum stellar mass. We ignore the methanol maser data
in case \ion{H}{ii} regions are detected because the radio continuum
data are more easily and accurately interpreted.

\subsection{Data on Water Masers}
One other alternative estimate can be obtained for the
$20~\rm{}km\,s^{-1}$ region. \citet{lu2015:20kms} present a detailed
and recent study of this region. Here we can use their detections of
water masers to obtain an alternative measure of the star formation
rate. Between 13 and 18 masers have line luminosities
$\gtrsim{}10^{-7}\,L_{\sun}$. We assume that every maser corresponds
to one star. These line luminosities imply bolometric luminosities
$\gtrsim{}50\,L_{\sun}$ \citep{palla1993:water-masers}, which
corresponds to stellar masses $\gtrsim{}2.75\,M_{\sun}$ (assuming the
mass--luminosity data from \citealt{mottram2011:timescales} can be
interpolated towards solar values using a power--law). One such star
requires a stellar group of 115 stars (Eq.~\ref{app-eq:n_cl}). The
full maser population thus implies between 1,480 and 2,050 embedded
stars, each with a mean stellar mass of $0.29\,M_{\sun}$. Water masers
are usually associated with the class~0 stage of star formation
\citep{furuya2001:water-maser} which has a duration of order 0.16~Myr
\citep{evans2009:c2d-summary}. Combination of these numbers gives a
star formation rate
$\sim{}3\times{}10^{-3}\,M_{\sun}\,\rm{}yr^{-3}$. This is
significantly in excess of the rate of
$\sim{}0.8\times{}10^{-3}\,M_{\sun}\,\rm{}yr^{-3}$ implied by the data
on \ion{H}{ii}~regions.

This estimate suffers from a number of uncertainties. First, it is not
clear that every luminous young star produces water masers. This could
mean that we miss a significant number of young stars. Second, a
single star might produce several masers. This could imply that we
overestimate the number of young embedded stars. Third, there is
significant scatter in the relationship between maser and bolometric
luminosities \citep{palla1993:water-masers}. Fourth, the current data
on the relationship between water masers and stellar evolutionary
stages is dated \citep{furuya2001:water-maser} and not necessarily
closely linked to the evolutionary stages of SF as understood and
defined today \citep{evans2009:c2d-summary}. Fifth, much of the
bolometric luminosity of lower--mass stars with water masers might
come from accretion. In that case the conversion from bolometric
luminosities overestimates stellar masses therefore the SF rate.
Table~\ref{tab:sf-tracers} lists the SF rate determined here, but
future updates appear likely.

\end{appendix}

\end{document}